\documentclass[useAMS,usenatbib,usegraphicx]{mn2e}

\newcommand{\mas}{$\mathcal{M}$}
\newcommand{\mastol}{$\mathcal{M/L}$}

\title[Mass of early-type galaxies]
{Stellar mass estimates in early-type galaxies: procedures, uncertainties and
models dependence}
\author[M. Longhetti et al.]{M. Longhetti$^{1}$\thanks{E-mail:
marcella.longhetti@brera.inaf.it}, P. Saracco$^{1}$\\
$^{1}$INAF - Osservatorio Astronomico di Brera, Via Brera 28, 20121 Milano
}
\begin{document}

\date{Accepted 2008. Received 2008; in original form 2008}
\pagerange{\pageref{firstpage}--\pageref{lastpage}} \pubyear{2008}
\maketitle

\label{firstpage}
\begin{abstract}
The aim of the present paper is to quantify the dependence of the estimates
of luminosities and stellar mass content of early-type galaxies
on the different models and model parameters which can be used
to analyze the observational data. The paper is organized in two parts. The first one analyzes
the dependence of the \mastol\ ratios and of the {\it k-}corrections in different bands
on model parameters (Initial Mass Function, metallicity, star formation history, age),
assuming some among the most popular spectrophotometric codes usually adopted to study
the evolutionary status of galaxies: Bruzual \& Charlot (2003, BC03), Charlot \& Bruzual (2008, CB08),
Maraston (2005, Ma05), Fioc \& Rocca-Volmerange (1997, PEGASE), Silva et al. (1998, GRASIL).
The second part of our work is dedicated to quantify the reliability and systematics affecting
 the mass and luminosity estimates obtained by means of the best fitting technique
applied to the photometric SEDs of early-type galaxies at $1<z<2$. 
To this end, we apply the best fitting technique 
to some mock catalogs built on the basis of a wide set of models of early-type galaxies.
We then compare the luminosity and the stellar mass estimated from the SED fitting 
with the true known input values. 
The goodness of the mass estimate is found to be dependent on the mass estimator
adopted to derive it, but masses cannot anyhow be retrieved better than within a factor 2-3,
depending on the quality of the available photometric data and/or on the distance of the galaxies
since more distant galaxies are fainter on average and thus affected by larger photometric errors.

Finally, we present a new empirical mass estimator based on the K band {\it apparent} magnitude and 
on the {\it observed} (V-K) colour. We show that the reliability of the stellar mass content 
derived with this new estimator for early-type galaxies and its stability 
are even higher than those achievable with the 
best classic estimators, with the not negligible advantage that it does not need any 
multi-wavelength data fitting.

\end{abstract}


\section{Introduction}
The formation and evolution history of the early-type galaxies represent a key
ingredient in the more global models of galaxy formation. 
Early-type galaxies are elliptical and bulge dominated galaxies which
contain most of the present stellar mass and which include the most
massive galaxies of the local Universe. They are characterized
by a pronounced 4000 \AA\ break, corresponding to rest frame red
UV-optical colours. This feature is typical
of evolved stellar populations (i.e. old with respect to the age of the Universe at
the redshift of the galaxy)  with almost absent star forming activity.

Two approaches can be pursued in the study of the past history
of this class of objects: the archaeological approach, that studies
in details local samples of galaxies attempting to derive their 
past history, or the direct analysis of samples of galaxies at
different redshift (i.e. at different stages of their evolution).
In both the two cases, the spectrophotometric models allow the
interpretation of the observed spectral properties of galaxies and thus they are a basic
ingredient to correctly derive their formation and evolution history.

One of the fundamental parameters used to derive the past 
history of the early-type galaxies is their stellar mass content. Indeed,
the measure of the stellar mass of early-type galaxies is involved in
some useful scale relations such as the {\it size-mass} relation by Shen et al. (2003).
Furthermore, the origin and/or evolution of the early-type galaxies seem to follow
different ways or at least different time scale as a function of their stellar
mass, the so-called {\it downsizing} (e.g., Bell et al. 2005; Thomas et al. 2005).
Thus, in the last years it has become important to assess the stellar mass content
of early-type galaxies in various range of redshift. 

The stellar mass content of galaxies is usually derived from luminosity measures
by means of conversion factors supplied by models
which for a given set of physical parameters well describe their observed Spectral Energy
Distribution (SED). The degeneracy among the model parameters which allow to
reproduce the observed SEDs of the galaxies (e.g. age, metallicity and star formation
history) and the differences in the conversion factors supplied by different models determine
the uncertainty of the mass estimates. 
The aim of the present paper is to study the dependence of the estimates of 
the stellar mass content and of luminosity of the early-type galaxies
on the choices of different available models and
of different parameters defining the models.
A forthcoming paper will focus on the
problem of the age determination.

The paper is organized in two parts. The first one (\S2 to \S3)
analyzes the dependence of the \mastol\ ratios and of the {\it k-}corrections in different bands
on model parameters (Initial Mass Function, metallicity, star formation history, age),
assuming some among the most popular spectrophotometric codes usually adopted to study
the evolutionary status of galaxies: Bruzual \& Charlot (2003, BC03), Charlot \& Bruzual (2008, CB08),
Maraston (2005, Ma05), Fioc \& Rocca-Volmerange (1997, PEGASE), Silva et al. (1998, GRASIL).
In particular,
\S2 introduces and defines the main parameters which can be varied
in the spectrophotometric models adopted to reproduce the spectral properties
of the early-type galaxies, and it presents the main features of the models which
will be used in the following sections, picking out the basic differences among them. 
\S3  presents the results which can be obtained
when different spectrophotometric codes are used to derive {\it k-}corrections and \mastol\ ratios
needed to transform the apparent magnitudes into mass estimates.
In the last part of this section, it is presented a self consistent relation
that allows to easily calculate \mas\ from the apparent K band magnitude as a function of {\it z}
that is valid for all the models here considered within 0.15 dex (a factor 1.4).

The second part of the paper (\S4) is dedicated to quantify the reliability and systematics which can
result in the mass and luminosity estimates obtained by means of the best fitting technique 
applied to the photometric SEDs of early-type galaxies
at $1<z<2$.  A wide range of simulated spectra are used to build mock photometric catalogs.
The measures of the luminosities and of the stellar masses of the simulated galaxies are compared
with the true known input values. Appendix A complements this comparison as far as the
\mastol\ ratios are concerned. In the end of this section, we present a test aimed at verifying
how different are the stellar mass estimates obtained
with different set of templates created with different codes (\S4.3).

Before concluding the paper, \S5 presents a new mass estimator
based on the {\it apparent} K band magnitude and (V-K) colour. It is demonstrated that
its reliability is similar to that of the best most commonly used mass estimators
despite the simplicity of its calculation that does not need any multi-wavelength
data fitting. \S6 summarizes the basic results found in the present work.

\section{The models}
In this section, we introduce the main parameters which can be varied
in the spectrophotometric models adopted to reproduce the spectral properties
of the early-type galaxies. The main features of the models which will be compared 
in the following sections are also presented.

\subsection{Parameters}
The spectrophotometric models are codes which produce the
expected Spectral Energy Distribution (SED) for a given set
of parameters describing a stellar population. The basic parameters
defining a stellar population are: 1) stellar metallicity $Z$; 2) Initial
Mass Function (IMF); 3) stellar evolutionary tracks; 4) stellar spectral library;
5) star formation history (SFH) and formation redshift ($z_{f}$) of the
stellar content.
Stellar evolutionary tracks and the stellar spectral library are usually
fixed or ``suggested'' by the spectrophotometric codes, and for those
compared in this work they will be listed below in the next part
of this section. As far as the other parameters are concerned (i.e. $Z$, IMF and SFH),
they represent the main adjustable variables when reproducing galaxies spectra.
In the following sections, we will present the different results which
can be achieved under different assumptions for these 3 parameters.

Since this work is devoted to the analysis of the spectral data
of early-type galaxies, few remarks on some restrictions on the
variability of SFH and $z_{f}$ have to be done.
Indeed, observations collected so far have outlined a picture in which
the bulk of stars in the early-type galaxies (at least in the most massive
among them, \mas$> 10^{11}$ \mas$_\odot$) formed at $z>2$ and passively evolved
down to the present epoch. For example,
from the observations of local and nearby samples of early-type
galaxies, it can be deduced that massive field early-type galaxies should have formed
their stellar content around $z\sim 2$ over short (i.e.  $\tau<1$ Gyr) star
formation timescales (Thomas et al. 2005).
Furthermore, the analysis of the Fundamental Plane of early-type
galaxies at $z\sim 1$ in the field (di Serego Alighieri et al. 2005,
van der Wel et al. 2005,
Treu et al. 2005, van der Wel et al. 2004, van Dokkum et al. 2003)
and in clusters (e.g. Bender et al. 1998) fully supports
a scenario in which their stars formed at $z > 2$ and only the less massive spheroids
could have been involved in star forming episodes at $z<1.2$.
Similar results come from the analysis of the Faber \& Jackson Relation
(FJR, Faber \& Jackson 1976) of a sample of field early-type
galaxies at $\sim0.7$ (Ziegler et al. 2005).
At higher redshift ($z>1$) confirmations of this picture are
recently coming out. For example, McCarthy et al. (2004) found a median
age of 1.2 Gyr in a sample of 20 evolved galaxies at $z \sim 1.5$,
and Daddi et al. (2005) report ages of about 1 Gyr for 7 galaxies
at $1.4 < z < 2.5$. Longhetti et al. (2005) from the analysis of a sample
of 10 massive (\mas$> 10^{11}$ \mas$_\odot$) early-type galaxies confirm that the
formation of early-type galaxies is well described assuming  $2<z_{f}<3.5-4$
and SF time scales shorter than 1 Gyr.
Thus, we consider as SFHs good to reproduce the spectral properties
of the early-type galaxies only those with SF time scale shorter than 1 Gyr.
We will also assume in the following an indicative formation redshift of their 
stellar content $z_{f}=4$ that will be used to associate to each value of 
$z$ the corresponding age of their stellar populations equal to
the time elapsed since their formation. 
Even in the possible case that the assembly history of early-type galaxies
was driven by major mergers at $z<2$, the latter must be ``dry'' (i.e. dissipationless,
Bell et al. 2006; van Dokkum 2005) and they should have small effects on
their global total star formation histories. On the other hand, if multiple frequent
minor mergers at $z<2$ contribute to grow the stellar mass of early-type galaxies,
these should add only small percentages to their total stellar content (Bournaud, Jog \& Combes 2007).
In both cases, the simplified assumption of the exponentially declining star formation history
with time scale shorter than 1 Gyr started at $z\sim4$ remains a realistic description of the 
formation and evolution history of the early-type galaxies at least as far as their
global photometric properties are concerned. Anyhow, all the results of the analysis
presented in the following sections are almost independent of the assumption 
of $z_{f}=4$, unless explicitly stated. 
A detailed analysis of the effects of varying $z_{f}$ can be found at 
{\it http://www.brera.inaf.it/utenti/marcella/}.

\subsection{Codes}
We have chosen to compare the output of some popular spectrophotometric
codes. Here the selected codes are listed and briefly described. 

\vskip 0.2 truecm
\noindent
{\bf Bruzual \& Charlot 2003 - BC03} \par\noindent
Among the possible choices offered by this code,
we select models based on the  stellar evolutionary tracks collected by the Padova 1994
library (i.e.  Alongi et al 1993; Bressan et al. 1993; Fagotto et al. 1994a,b;
Girardi et al. 1996), and which include the stellar theoretical spectral library 
BaSeL (3.1) (see Bruzual \& Charlot 2003 for details) that provides
a spectral resolution R=300 over the whole spectral range from
91\AA\ to 160 $\mu$m. Models of stars with $T > 50,000$ K are taken from Rauch (2003),
and the spectral models of the Thermally-Pulsating Asymptotic Giant Branch (TP-AGB) phase
are based on Vassiliadis \& Wood (1993).
 We consider models at four different metallicities:
$Z_\odot$, 0.2$Z_\odot$, 0.4$Z_\odot$, and 2.5$Z_\odot$. Within this code,  we also consider different
IMFs, all assuming 0.1M$_\odot<\mathcal{M}<100$M$_\odot$: Salpeter (Sal; Salpeter 1955),
Kroupa (Kro; Kroupa 2001), Chabrier (Cha; Chabrier 2003).
The SF histories include exponentially declining SFR with $\tau= 0.1, 0.4, 0.6, 0.8, 1.0$ Gyr.

A new version of this code, that includes the new  prescription of Marigo \& Girardi (2007)
for the TP-AGB evolution of low- and intermediate-mass stars,
is going to be published soon. We consider its preliminary version in
the following comparisons, referring to it as the {\bf CB08} code (Charlot \& Bruzual 2008).
Within this code, we consider four different metallicities (the same as in BC03)
and the Chabrier IMF.

\vskip 0.2 truecm
\noindent
{\bf Fioc \& Rocca-Volmerange 1997 - PEGASE} \par\noindent
The code by Fioc \& Rocca-Volmerange (1997) is based on the same tracks
and the same stellar spectral library as BC03. The only difference between
the two codes is given by different stellar spectra assumption for the hottest stars
(T$>50,000$ K) and different prescription for the TP-AGB phase, 
both affecting the results at young ages ($\le 10^{8}$ yr). 
Indeed, spectra of hot stars are taken
from Clegg \& Middlemass (1987), while the models of the TP-AGB phase are 
based on the prescriptions of Groenewegen \& de Jong (1993).
For comparison with the results obtained with the BC03 code
and with the other codes described in the following, we have built
models with exponentially declining SFR with time scale $\tau = 0.1, 0.4, 0.6, 
0.8, 1.0 $ Gyr and Salpeter IMF (assuming 0.1M$_\odot<\mathcal{M}<120$M$_\odot$).
As far as the metallicity is concerned,  consistent chemical evolution
is included in the PEGASE models which start from metallicity $Z=0$
at time $t=0$. In the comparison made in the following sections
among the results obtained with different models, we should keep in mind that PEGASE models are
characterized by low values of the stellar metallicity at young ages (e.g. at
1 Gyr $Z\simeq 0.5 Z_{\odot}$) while the maximum average value reached (at ages greater than
10 Gyr) is 75\% of the solar value.
No infall has been selected among the parameters
(i.e.  all the gas available to form stars is assumed to be in place
at time $t=0$). No galactic wind has been selected and the default
value of 0.05 has been assumed for the parameter representing
the fraction of close binary systems. Extinction has been excluded
while the calculation of the emission lines expected in
the resulting spectra of this set of models has been included.

\vskip 0.2truecm
\noindent
{\bf Silva et al. 1998 - GRASIL} \par\noindent
GRASIL is a code to compute the spectral evolution of stellar systems
that takes into account the effects of dust during the star forming phases
of their evolutions. Evolutionary tracks are selected from
the Padova 1994 library as in BC03 and PEGASE. The stellar spectral
library is based on the Kurucz (1992) models and is only partly
coincident with the BaSel library adopted by PEGASE and BC03.
Hot stars spectra are modelled  with blackbody spectra.
The TP-AGB phase  is described on the basis of the prescription of
Vassiliadis \& Wood (1993), as in the BC03 code.
The parameters selected by the authors to describe 
their {\it giant elliptical galaxy} discussed in Silva et al. (1998)
and listed in their table 1 and table 2 have been assumed.
The SFR is described in the framework of an infall model by the Schmidt law
and it is stopped after
1 Gyr, and the assumed IMF is the Salpeter one (assuming 0.15M$_\odot<\mathcal{M}<120$M$_\odot$), 
so that the obtained model from this point of view
is comparable with those created with the SFHs adopted
in the other codes.
All the technical parameters regulating all the physical
processes included in this code have been kept to their
default numbers. It is important to note that also in this case, as in the
case of PEGASE models, metallicity increases with time, and it reaches the maximum
average value of 85\% of the solar one.

\vskip 0.2truecm
\noindent
{\bf Maraston 2005 - Ma05} \par\noindent
The code by Maraston (2005) is quite different from the previous ones on
many aspects. First of all, it builds the expected SEDs of stellar populations
on the basis of the `fuel-consumption' approach while BC03, PEGASE and GRASIL adopt
the approach of `isochrone synthesis' (see details in Chiosi et al. 1998; Charlot \& Bruzual 1991).
Furthermore, Ma05 code is based on the stellar evolutionary tracks 
by Cassisi et al. (1997, 2000), Cassisi \& Solaris (1997). 
Finally, the main difference between this code and the previous ones 
is the different treatment of the TP-AGB phase. Indeed, 
the prescriptions of the previous codes
calibrate the duration of the TP-AGB phase, while the prescription of Ma05 calibrates the
fractional luminosity contribution to the total bolometric light.
The resulting spectra, even if both the two kinds of calibrations are based
on the same set of observed data (Frogel et al.  1990), include a larger contribution of the TP-AGB phase
in Ma05 than in the other codes (see also Maraston et al. 2006), with the exception of CB08 code that
 includes a new AGB phase description by Marigo \& Girardi (2007) with a fractional
luminosity contribution more similar to that adopted by Ma05.

A part from the important differences reported above, the
stellar spectral library adopted by Ma05
is the same as the previous codes (i.e., BaSel library) for stars at $T < 50,000$ K, while
for the hottest stars blackbody spectra as in the GRASIL code are adopted.
The composite stellar population spectra available from
the author's web page include solar metallicity models
obtained with exponentially declining SFR with time
scale $\tau =0.1, 0.4, 1.0$ Gyr. They have been obtained assuming the
Salpter IMF.

\vskip 0.2 truecm
Before going into the details
of the comparisons between the results obtained with the different codes in
the analysis of the early-type galaxies, it is worth noting that the models
by Ma05 for ages between $\sim 0.2$ and $\sim 2$ Gyr are expected
brighter and redder than those obtained with the other codes (Maraston et al. 2006). 
This is due to the
different description of the full AGB phase and to the different isochrones 
adopted in Ma05 code with respect to the other ones.

\section{From luminosity to stellar mass}

The stellar mass of a galaxy is directly linked to its luminosity,
so that the most obvious way to derive this important parameter
is to convert a luminosity measure into a stellar mass estimate.
In this section, we will analyze the details of this conversion,
and its dependence on models and model parameters, assuming that 
redshift and age are known and fixed.

\subsection{Stellar mass: definitions}
Before going on, 
the concept of stellar mass itself needs to be clearly stated. 
Indeed, there are three different way to define the mass of a 
galaxy (see Renzini 2006) when it is related to the output of models:

\begin{enumerate}
\item the mass of gas burned into stars from the epoch of its formation to 
the time corresponding to its age,
resulting from the integration over time of its SFR: 

\begin{equation}
\mathcal{M}_{SFR}(t)=\int_0^tdt'SFR(t')
\end{equation}

\noindent
(in BC03 models the value of the mass above defined can be read
in column 9 of files {\it .4color}).

\item the mass contained at any epoch into stars, both those
still surviving and those which are dead remnants: 

\begin{equation}
\mathcal{M}_{star}(t)=\int_0^tdt'SFR(t')-\int_0^tdt'\mathcal{M}_{loss}(t')
\end{equation}
where $\mathcal{M}_{loss}(t')$ is the mass of gas returned to the interstellar
medium from stars at any time $t'$
(in BC03 models the value of this difference is calculated and reported in
column 7 of files {\it .4color}).

\item the mass that at any epoch is contained into still surviving stars:

\end{enumerate}

\begin{equation}
\mathcal{M}_{alive}(t)=\int_0^tdt'SFR(t')-\int_0^tdt'\mathcal{M}_{loss}(t')-\mathcal{M}_{r}(t)
\end{equation}
where $\mathcal{M}_{r}(t)$ is the mass of dead remnants at the time $t$ 
(in BC03 models it needs to be calculated from the above formula where
$\mathcal{M}_{r}(t)$ is read in column 12 of files {\it .3color}).

\begin{figure}
\centering
\includegraphics[width=9.5truecm]{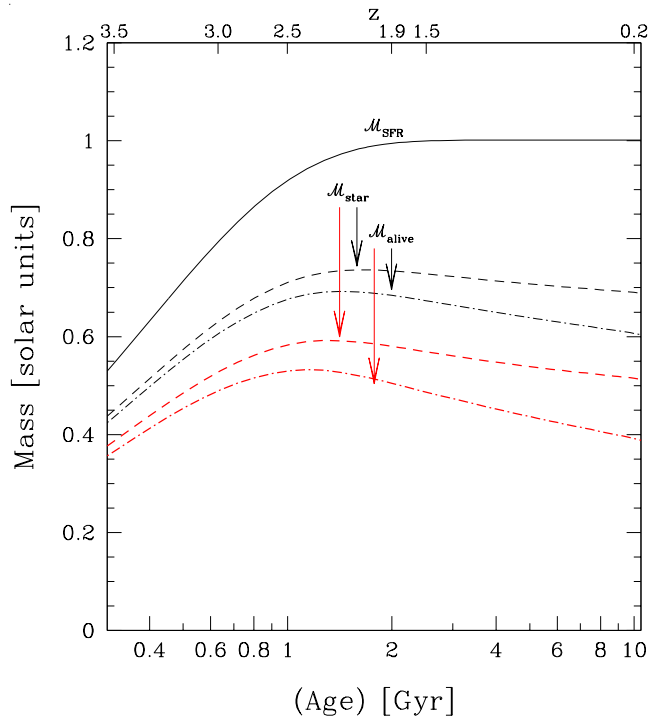}
\caption{The mass content of a simulated galaxy is plotted as function of
its age (bottom axis) and $z$ (assuming $z_{f}=4$, top axis).
The black continuum line stands for the mass obtained by integration
over the age of the model of a declining SFR with $\tau = 0.4$ Gyr,
normalized to the final total mass
(in BC03 models, in column 9 of files {\it .4color}).
Short dashed lines stand for the stellar mass still kept into
stars at any age (i.e. it takes into account the stellar mass loss
of some evolutionary phases, in BC03 models, in column 7 of files {\it .4color}).
Dotted - short dashed lines stand for
mass still kept into surviving stars alone (mass of dead remnants -e.g. in
column 12 of files {\it .3color} in BC03 models - is subtracted).
Black lines refer to models built with the Salpeter IMF,
while red (grey) lines represent models obtained with Chabrier IMF
(both assuming solar metallicity).
}
\end{figure}

\noindent
While \mas$_{SFR}$ is the same for different models since it depends only
on the selected SF history, \mas$_{star}$
depends on the assumed IMF within the code, and on the
stellar evolution picture on which the code itself is based.
Figure 1 shows how different are the values corresponding to the three
definitions of mass for the same model of galaxy 
and their dependence on the assumed IMF (Salpeter and Chabrier), in the case
of a declining SFR with $\tau = 0.4$ Gyr.
For ages greater than few Gyr the mass still locked into
stars is about 67\% (50\%) of the total mass involved in the past
star formation activity, and only the 57\% (35\%) is that still locked into
surviving stars for Salpeter (Chabrier) IMF.
Smaller differences are produced by the assumption of different
stellar tracks. Indeed, the same model reported above but built with the
code Ma05 (replacing BC03 code) and Salpeter IMF has 72\% of the total mass
involved into past star formation activity locked into stars
at ages of about 12-15 Gyr, and 61\% locked into still alive
stars, to be compared with 67\% and 57\% respectively as
resulting from BC03 code.

In the following, we always refer to the total mass of the models expressed
in equation (2), \mas=\mas$_{star}$, that is the most common definition adopted in
recent works on stellar mass determinations (e.g.,
Fontana et al. 2006, van der Wel et al. 2006,  Pozzetti et al. 2007).
Since in the case of the SF histories which reasonably describe that of the early-type galaxies
at ages greater than 1 Gyr \mas$_{SFR}$ is almost constant to its maximum value,
we propose here some useful fits which allow to transform
\mas$_{star}$ into \mas$_{SFR}$ for different IMFs and as a function of
age (expressed in Gyr) and of {\it z} (assuming as formation redshift $z_{f}=4$) which
are valid for exponentially declining SF histories with time scale $\tau \le 0.8$ Gyr:

$$\mathcal{M}_{SFR}={\mathcal{M}_{star} \over [c_{0}^{age}-(0.01 \times age)]}$$
$$\mathcal{M}_{SFR}={\mathcal{M}_{star} \over [c_{0}^{z}+(0.05 \times z)]}$$

\noindent
where $c_{0}^{age}$ is 0.75 assuming the Salpeter IMF and 0.65 for Kroupa and Chabrier IMFs,
while $c_{0}^{z}$ is 0.62 for Salpeter IMF and 0.51 for Kroupa and Chabrier IMFs.
The accuracy of the above equations is better than 10\%, even if $z_{f}>4$.

\subsection{Stellar mass: dependence on the model parameters}

The conversion of a luminosity measure into a stellar mass estimate
can be summarized by the following equation:

\begin{equation}
\mathcal{M}_{gal} [\mathcal{M}_{\odot}] = \mathcal{L}_{\lambda}^{gal} \ \mathcal{M/L}_{\lambda}
\end{equation}
where \mas$_{gal}$ is the stellar mass of the galaxy expressed in solar masses,
$\mathcal{L}_{\lambda}^{gal}$ is the
luminosity of the galaxy at $\lambda$ in solar luminosities, and $\mathcal{M/L}_{\lambda}$
is the mass to light ratio at $\lambda$ in solar units. The luminosity of the galaxy
$\mathcal{L}_{\lambda}^{gal}$  can be calculated from:

\begin{equation}
\mathcal{L}_{\lambda}^{gal} [\mathcal{L}_{\odot}] = 10^{[-0.4  (M_{\lambda}^{gal} -\  M_{\lambda}^{sun})]}
\end{equation}
where $M_{\lambda}$ stands for the absolute magnitude at $\lambda$.
If we merge the equations (4) and (5), and we consider that $M_{\lambda}^{gal}$ is
derived from the apparent magnitude $m_{\lambda'}^{gal}$, we obtain:

$$\log(\mathcal{M}_{gal}) = \log(\mathcal{M/L}_{\lambda})+0.4kcor_{\lambda}+2\log(d_{pc})+ \ \ \ \ \ \ \ \ \ \ \ \ $$

\begin{equation}
\hfill -2.0+0.4 M_{\lambda}^{sun}-0.4m_{\lambda'} \ \ \ 
\end{equation}

\noindent
where $d_{pc}$ is the distance expressed in $pc$ that depends from the assumed
cosmology and it is a function of $z$, $M_{\lambda}^{sun}$ is the absolute magnitude of the sun in
the chosen filter centered at $\lambda$, while $m_{\lambda'}$ is the apparent
magnitude of the galaxy in the observed filter centered at $\lambda'$.
In Table 1 the most common values of $M_{\lambda}^{sun}$ for
optical, near-UV and near-IR bands are given.

\begin{table}
\caption{Absolute magnitude of the Sun.}
\centerline{
\begin{tabular}{ccc}
\hline
 band & $M_{\lambda}^{sun}$ & {\it Note} \\
\hline
 U & 5.61 & a \\
 B & 5.48 & a \\
 V & 4.83 & a \\
 R & 4.31 & a \\
 I & 4.02 & a \\
 J & 3.77 & b \\
 H & 3.45 & b \\
 K & 3.41 & a \\
\hline
\end{tabular}
}
{\it Note}: (a) from the solar spectrum available at the CALSPEC
database at STScI (Allen 1976);
(b) values derived from $M_{K}^{sun}$ by means of colors
calculated on the Kurucz synthetic spectrum G2 V.
\end{table}

\begin{figure*}
\centering
\includegraphics[width=15.8truecm]{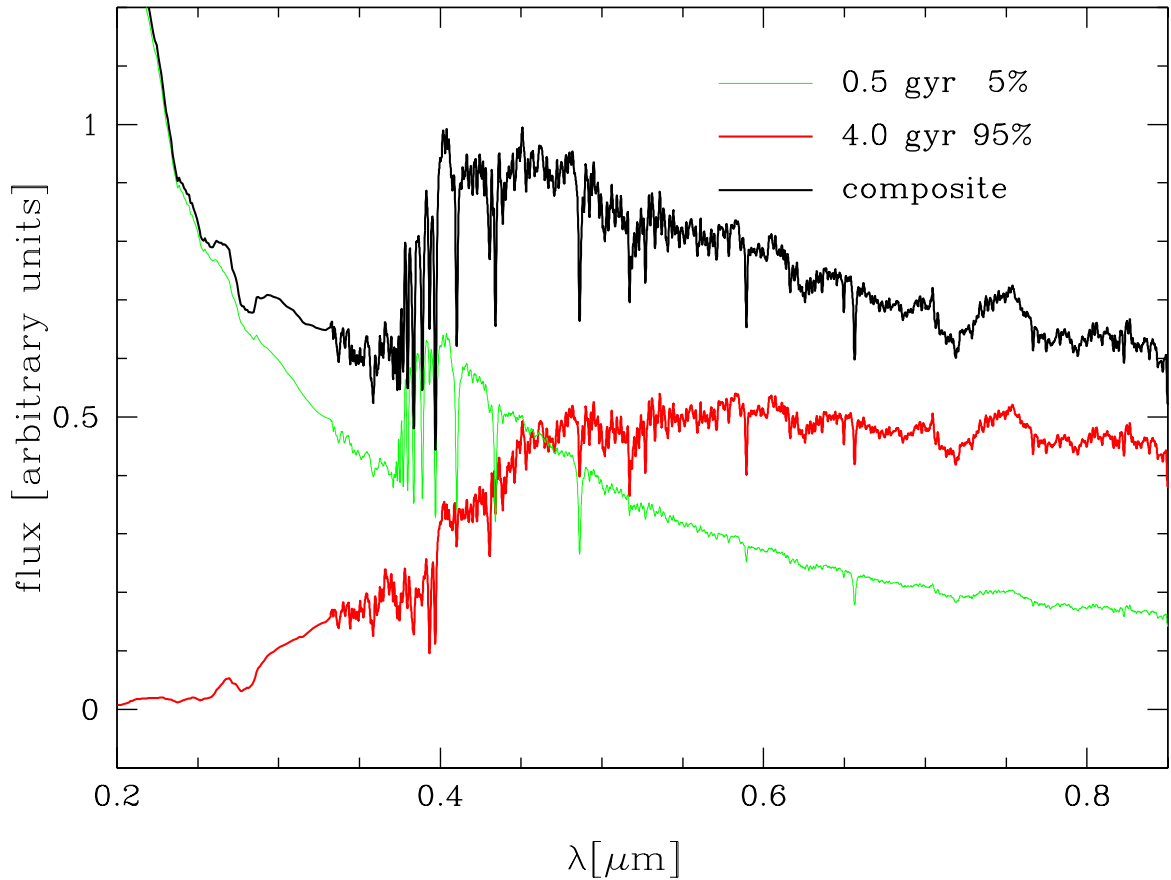}
\caption{Thick black line: synthetic spectrum of a composite
stellar population obtained by summing 95\% of the mass
of a 4 Gyr old population (thick grey/red line) and 5\% of the mass
of a 0.5 Gyr stellar component (thin grey/green line).
Both the stellar populations correspond to solar metallicity and to
the Salpeter IMF, and they have been derived from the exponentially
declining star formation history with $\tau=0.4$ Gyr.  The contribution to the total
luminosity of the young SSP is more than 90\% at
$\lambda < 0.3 \mu m$, and similar to the contribution
of the older much more massive component at
$0.3\mu m < \lambda < 0.5 \mu m$. At $\lambda > 0.5 \mu m$
the contribution to the total luminosity of the young component
is decreased down to less than 30\%.
}
\end{figure*}

Spectrophotometric models are involved in the determination of two quantities of 
equation (6): {\it i)} the conversion factor \mastol\ 
and {\it ii)}  the {\it k-}correction needed to calculate the luminosity
of the galaxy. In the following subsections, it will be demonstrated that
the greatest accuracy in the determination of both the \mastol\ ratio and the luminosity
of galaxies is achieved in the near-IR bands. 
Indeed, optical blue and UV ranges
(i.e. $\lambda < 5000$ \AA) are typically dominated by
very luminous young stellar populations even when they represent few percentages
of the total stellar content of the galaxies (see figure 2).
In other words, the total stellar mass content is better traced by the near-IR luminosities
because it does not depend on many details of the SF history followed by the galaxies.
Moreover, the continuum shape of the near-IR spectral region is less
dependent on the age of the stellar populations (see figure 2) with respect
to the UV region, and this is why {\it k-}corrections (i.e. luminosities) can be determined with
larger precision in the K band than in the V one.

\subsubsection{Mass to light ratio: useful relations}

The conversion factor between luminosities and stellar masses provided by 
the models (\mastol) depends first of all on the chosen wavelength band.
Figures 3 and 4 show the value of \mastol\ as a function
of age and for different spectrophotometric codes, metallicities and SF histories,
in the V and K band, respectively.

\begin{figure}
\centering
\includegraphics[width=8.7truecm]{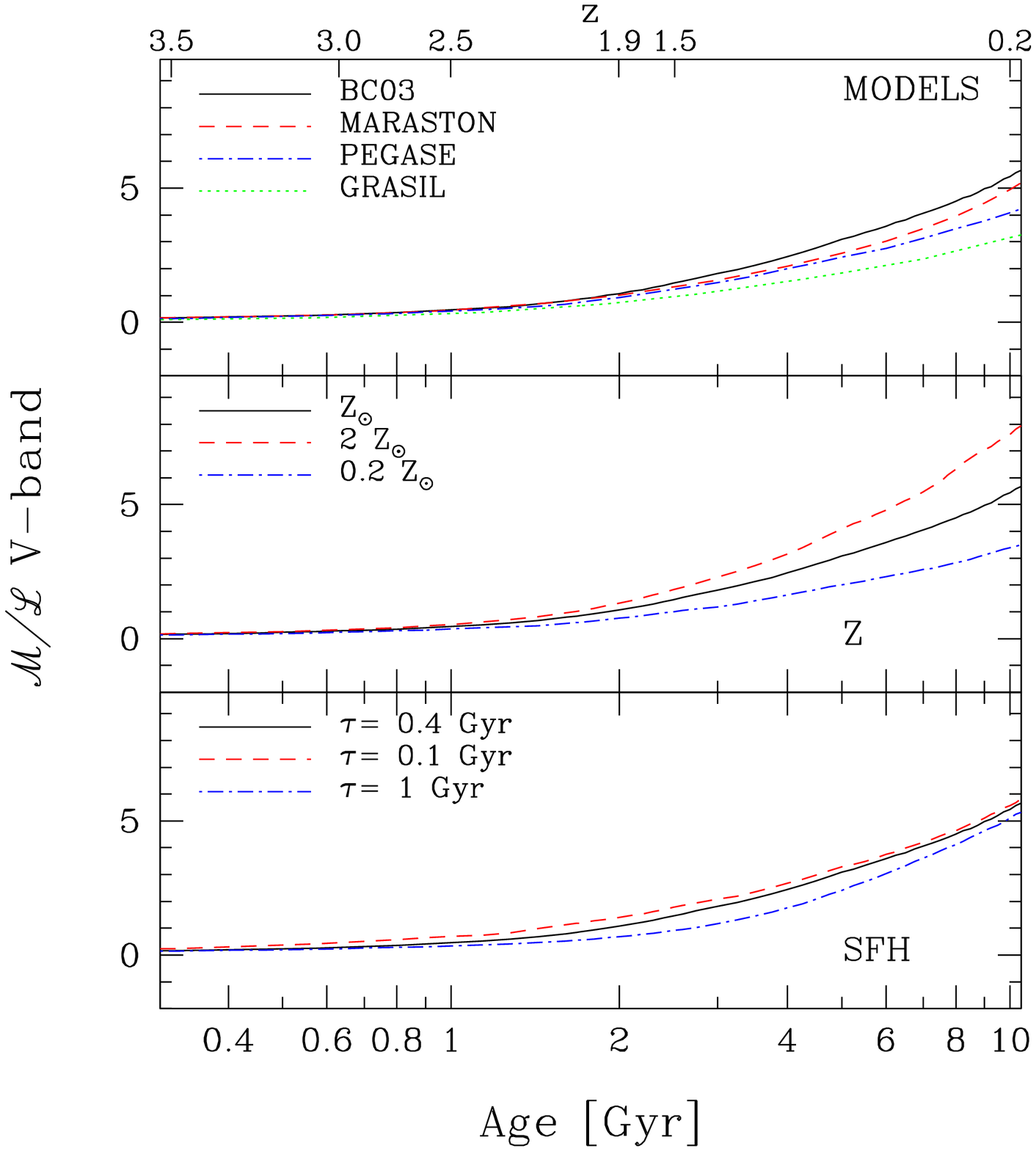}
 \caption{\mastol\ ratio in the V band as a function of age (bottom axis) and $z$ (assuming $z_{f}=4$, top axis).
In the {\bf top panel}, results obtained with different codes for the same model galaxy are presented.
The model galaxy has been built following an exponentially declining star formation rate with
time scale $\tau=0.4$ Gyr at solar metallicity (Salpeter IMF), with the exception of the GRASIL code
for which the Silva et al. (1998) {\it reference} model for elliptical galaxies has been considered (see
details in \S2.2).  In the {\bf middle panel}, the results obtained at three different values
of metallicity are displayed. Models have been built adopting the BC03 code and the exponentially 
declining star formation rate ($\tau=0.4$ and Salpeter IMF). The {\bf bottom panel} presents
the results obtained assuming three different time scales of the declining star formation rate,
for models built adopting the BC03 code at solar metallicity (Salpeter IMF).}
\end{figure}

\begin{figure}
\centering
\includegraphics[width=8.7truecm]{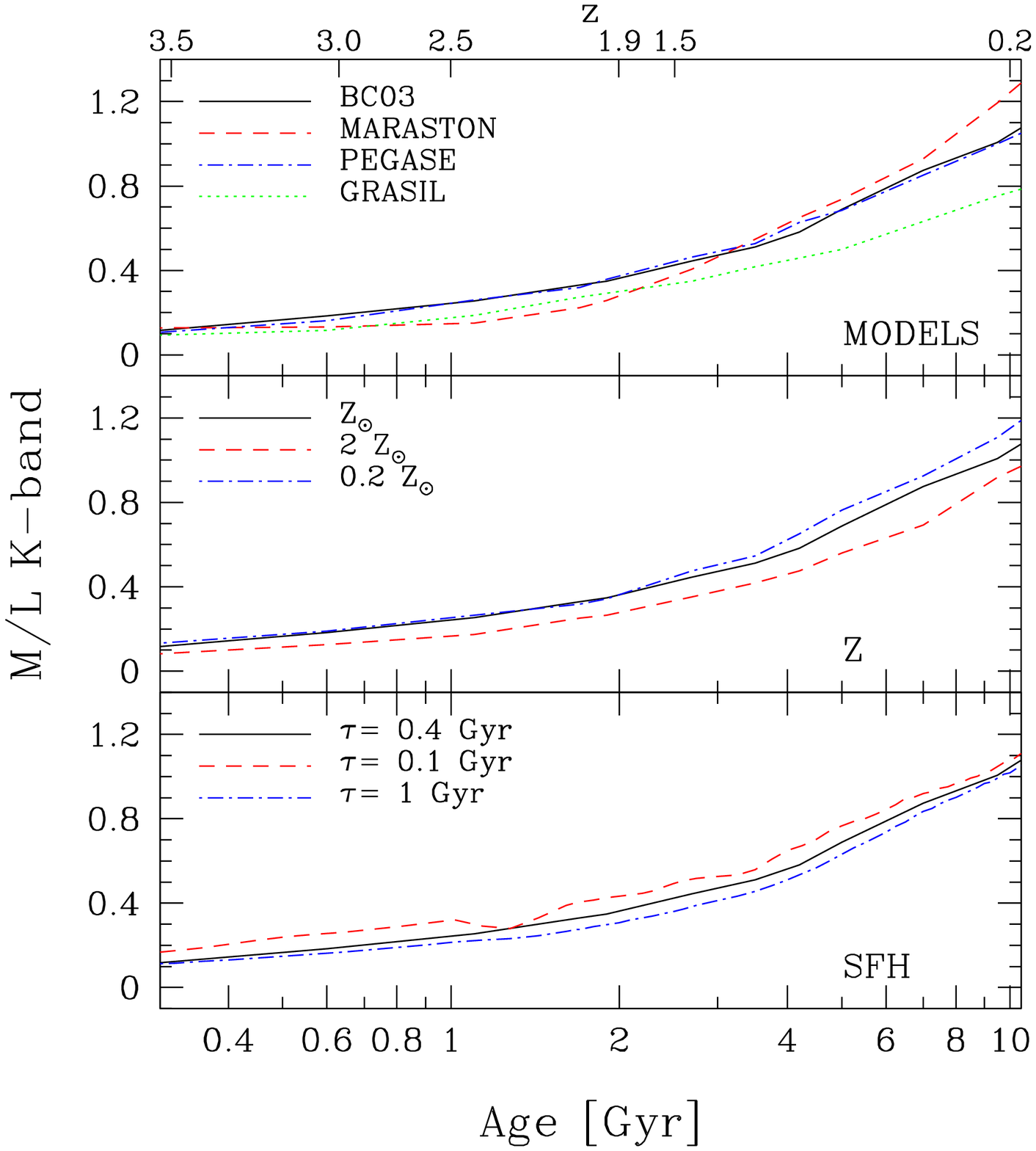}
 \caption{\mastol\ ratio in the K band as a function of age (bottom axis) and $z$ (assuming $z_{f}=4$, top axis).
The three panels show the effect of varying the code used to model the galaxies, the metallicity and
the star formation history as in figure 3.}
\end{figure}

\noindent
In both the two bands the largest variability
of \mastol\ is given by age variations, but
what it should be noticed from the comparison of the two figures is the different
y-axis scale that reveals a variation of \mastol\ in the V-band
of an order of magnitude against 50\% of variation as a maximum
of \mastol\ in the K-band. In other words, small errors in the age assumed to reproduce
the galaxies produce large errors in the resulting \mastol\ in the optical
bands, while have almost negligible effects in the determination of the \mastol\ 
in the near-IR bands. For example, at solar metallicity, a standard early-type galaxy is
characterized by $\Delta(\mathcal{M/L}_{V})/\Delta(age)=1.3$ and
$\Delta$(\mastol$_{K}$)/$\Delta(age)=0.2$ if age is expressed in Gyr. This means
that an uncertainty of 1 Gyr in the age determination of a 4 Gyr old galaxy
 leads to an uncertainty in the \mastol\ value of
0.2 in the K-band and of 1.3 in the V-band.
For the above reason, it is strongly suggested to use near-IR bands luminosities
when reliable mass estimates have to be derived.

Figures 4 and 5 contain all the dependencies of K-band \mastol\ ratio on models
and on their parameters. In particular, from the upper panel of figure 4
we can appreciate the small dependence of \mastol\ 
on the spectrophotometric code (and implicitly 
on the possible different assumptions of the stellar tracks).
Largest discrepancies are found between values obtained from Ma05 models
with respect to BC03 ones. Indeed, K band \mastol\ ratios from Ma05 models 
at ages lower than 2 Gyr reach values which are only 60\% of those obtained with the BC03 code.
This was expected since Ma05 models are brighter 
than the other models at ages between $\sim0.2$ and $\sim2$ Gyr, due
to their different prescriptions of the TP-AGB phase and different
isochrones adopted in the code (see \S2.2).
At the same time at old ages the values of \mastol\ found by means of the Ma05
code are higher of a factor 1.2 than those found with the BC03 code.
At old ages, large discrepancies are found also
between the values of the K-band \mastol\  ratios
obtained with the GRASIL code with respect to the BC03 one. Indeed, the GRASIL values differ
from the BC03 ones of a factor 0.8 at ages older than 7 Gyr, i.e.  $z<0.5$.
On the contrary, a better agreement is found between the PEGASE values and the BC03 ones
over the whole range of ages/redshift.

The middle panel of figure 4 shows the dependence of \mastol$_{K}$ on the
stellar metallicity. The maximum difference between the values obtained
at different metallicities is encountered at ages larger than 2 Gyr, 
when \mastol$_{K}^{Z_{\odot}}$ is about 90\% of  \mastol$_{K}^{0.2Z_{\odot}}$,
and 1.2 times \mastol$_{K}^{2Z_{\odot}}$. As an example, between $z=1$ and $z=2$,
assuming $z_{f}=4$, \mastol$_{K}$ varies from 0.6 to 0.3 for the effect of age variations
while it varies from 0.5 to 0.7 (0.2 to 0.3) for metallicity variations at $z=1$ ($z=2$).
Thus, at least in the ages/redshift range on which this work is focused,
metallicity effects are of a second order with respect to the 
age effects.

The bottom panel of figure 4 confirms also a small dependence of \mastol\  
 on the SF history selected to model the galaxy, at least for
SF time scales smaller than 1 Gyr for which the variation of \mastol\ 
is about 30\% as a maximum. 
Figure 5 explores the dependence
of \mastol\  on the IMF for three different SF histories. It clearly
appears that this parameter of the models is the main variable
that produces uncertainty in the resulting value
of \mastol. 
At the same time,
the \mastol\  values obtained within different IMFs
have almost fixed ratios, so that we can consider different
IMFs as different scaling factors for stellar mass calculations.
In figure 5 we also report values obtained with the CB08 code and Chabrier IMF,
which are systematically lower than those obtained with the same IMF
but with the BC03 code. Indeed, this was expected, at least at young ages
(i.e. $\sim 1-2$ Gyr) where the different treatment of the TP-AGB phase
produces an increase of the luminosity with respect to the one derived with the BC03 code.
Anyhow, we remember that the CB08 code is diffused in a preliminary version,
and we wait for the authors publication on the details of the new code to more precisely understand
this difference.

A complete summary of the results obtained with various
models in the calculation of  \mastol$_{K}$ as
a function of age and $z$ 
can be found at {\it http://www.brera.inaf.it/utenti/marcella/}.
In particular, since \mastol\ ratios in the near-IR bands are quite homogeneous 
and only slightly dependent on
codes and details of the models used to calculate them, at the above address  we suggest 
a set of {\it reference} values of \mastol$_{K}$ as a function of $z$  and of
the age of the galaxy, besides  the parameters needed to transform $\mathcal{M}_{star}/\mathcal{L}$
into $\mathcal{M}_{SFR}/\mathcal{L}$ in case of exponentially declining SF history with
time scale $\tau=0.4$ Gyr.
In \S3.3 we supply with an easy use of all those  numbers, proposing the fit
of \mastol$_{K}$ as a function of age and of $z$ (assuming $z_{f}=4$).

\begin{figure}
\centering
\includegraphics[width=8.7truecm]{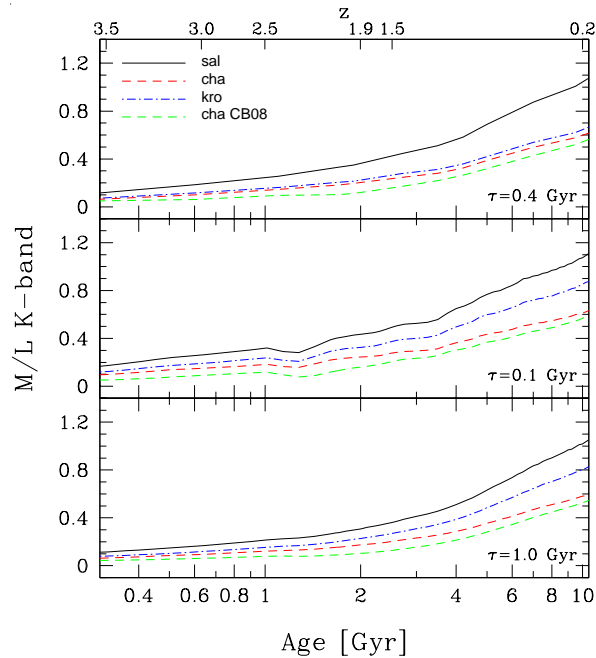}
\caption{\mastol\  ratio in the K band as a function of age (bottom axis) and $z$ (assuming $z_{f}=4$, top axis).
The values obtained assuming different IMFs in the BC03 models (solar metallicity) are shown for exponentially
declining models with $\tau=0.4$ Gyr (top panel), $\tau=0.1$ Gyr (middle panel) and $\tau=1.0$ Gyr (bottom panel).
For comparison, results obtained assuming the Chabrier IMF but with the new version of the BC03 code (i.e. CB08)
are displayed in all the three panels.}
\end{figure}

\subsubsection{{\it k-}corrections: useful relations}
{\it k-}corrections are needed to transform apparent magnitudes in the observed filter
at $\lambda'$ into absolute magnitudes in the same or another filter at
$\lambda$:

$$kcor=-2.5\log { {\int{F_{z}(\lambda)S_{1}(\lambda)d\lambda}\over\int{F_{0}(\lambda)S_{2}(\lambda)d\lambda}}
 {\int{S_{2}(\lambda)d\lambda}  \over \int{S_{1}(\lambda)d\lambda}} } \ \ \ \ \ \ \ \ \ \ \ \ \ \ \ \ \ \ \  $$

\begin{equation}
\ \ \  \ \ \ \ \ \ \ \ \ \ \ \ \ \ \ \ \ \ \ +2.5\log(1+z)+(ZP_{S_{1}}-ZP_{S_{2}})
\end{equation}

\noindent
where $F_{z}(\lambda)=F_{0}(\lambda{1\over 1+z})$, S1 and S2 represent the observed and reference filters respectively, 
and ZP$_{i}$ is the zero point for the filter ${i}$.
In the definition of the {\it classic k-}correction 
S1 and S2 are coincident, and the definition can be simplified as:

\begin{equation}
kcor=-2.5\log { {\int{F_{z}(\lambda)S(\lambda)d\lambda}\over\int{F_{0}(\lambda)S(\lambda)d\lambda}} } +2.5\log(1+z)
\end{equation}

\noindent
while that defined in equation (7) is sometimes called {\it colour k-}correction.

Figures 6 and 7 show the classic {\it k}-corrections in the K and V bands respectively, as a 
function of age and for different models and models assumptions.
The two figures do not display the effect obtained by varying the IMFs within the codes,
because it has been found to be completely negligible.
Figures 6 shows the high dependence of the near-IR {\it k}-corrections on
stellar metallicity that is comparable to its dependence on age, while
different SF histories display similar values. On the other hand,
figure 7 makes evident that in the optical bands age remains the main variable
driving the determination of the {\it k}-corrections (see the y-axis scale), 
while metallicity effects
can be neglected. Moreover, there appears to be a  large influence
of even small variations in the stellar compositions of model galaxies, that we have displayed 
as dependence on the SF history.  The absence of young components
at age $<4$ Gyr for short SF time scale (e.g., $\tau=0.1$ Gyr) produces values which 
largely differ with the V band {\it k-}corrections calculated with models based 
on longer SF time scales.

As far as the different codes are concerned, we find again a different
behavior of the {\it k-}corrections in the K band with respect to the V band.
Indeed, the near-IR {\it k-}corrections values agree quite well among the different
codes here analyzed, with the only exception of the Ma05 and CB08 models.
In particular, PEGASE models return on average the same value of BC03 models within
0.96-1.17. On the other hand, the GRASIL models give
K band {\it k-}corrections systematically smaller than the BC03 ones, but
still within less than 20\% (i.e., between 0.85 and 0.93).
The Ma05 models on the contrary supply K band {\it k-}corrections values
which, while at ages older than 3 Gyr (i.e. $z<1.5$) are larger than BC03 ones
(within a factor 1.17), at young ages they become much smaller down to 50\%.
Even larger discrepancies are found between BC03 models and CB08 ones, giving the latter
systematically smaller K band {\it k-}corrections down to 30\% of the former.
The latter two models are those which include more refined prescriptions 
for the TP-AGB phase which influence the continuum shape at young ages.

In the V band, the models which produce much different {\it k-}corrections values
with respect to the others (up to 4 times the values of the other models) 
are those built with the GRASIL code. In fact,
this reflects the different SF history that the GRASIL models assume
with respect to the exponentially declining star formation rates
with $\tau=$0.4 Gyr common to all the other models. Indeed, we have
already noticed the large dependence of optical {\it k-}corrections on the SF history of the 
models. As far as the other models are concerned, they generally agree quite well
with each other, with larger discrepancies found at intermediate ages (i.e. 2-5 Gyr, within a factor 1.3).

Summarizing, it can be said that the determination of the classic {\it k-}corrections 
in the optical bands present
larger uncertainties because of their large variability ($\sim 5$ magnitudes) as a function of age, and 
of their large dependence on the assumed SF history. On the other hand, the near-IR bands {\it k-}corrections
can be determined with larger accuracy because their variability is kept below 1.5 magnitudes
for whatever parameter or model is varied. Their largest uncertainty is due to metallicity variations
and to the choice of the code adopted to model galaxies between BC03, PEGASE and GRASIl codes on one hand
and Ma05 and CB08 codes on the other one.

\begin{figure}
\centering
\includegraphics[width=8.7truecm]{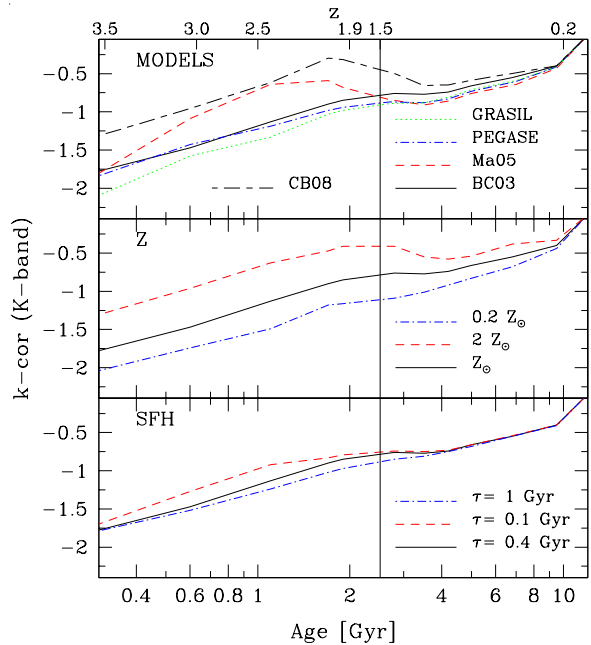}
 \caption{{\it k-}correction in the K-band as a function of age (bottom axis) and $z$ (assuming $z_{f}=4$, top axis).
Panels display the effects of varying models, metallicity and star formation history as in figure 3.}
\end{figure}

\begin{figure}
\centering
\includegraphics[width=8.7truecm]{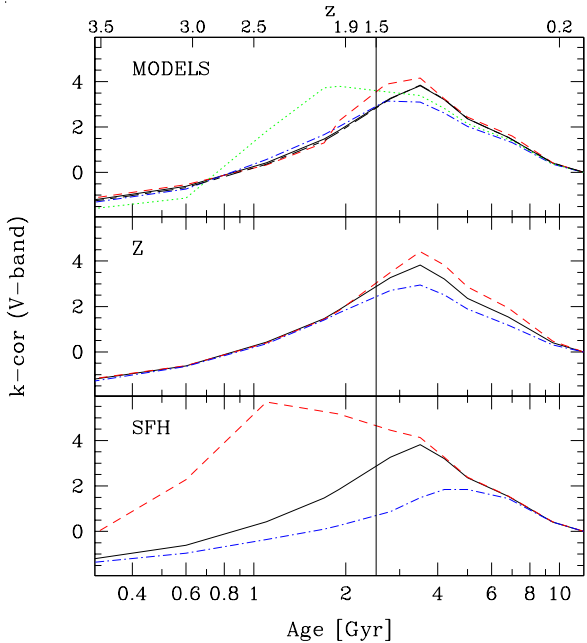}
 \caption{{\it k-}correction in the V-band as a function of age (bottom axis) and $z$ (assuming $z_{f}=4$, top axis).
Panels display the effects of varying models, metallicity and star formation history as in figure 3}
\end{figure}

Figures 8 and 9 present the colour {\it k-}corrections in the K and V band, respectively.
Colour {\it k-}corrections, defined as in equation (7), minimize the dependence on models
and model parameters assuming a starting observed band as clos as possible to the 
rest frame band in which the absolute magnitude has to be calculated.
In particular,  the K band {\it k-}corrections have been derived
from the observed bands from K to Spitzer 8$\mu$m, while the V band ones
from V to K observed bands.

A complete summary of the results obtained with the different
models in the calculation of the {\it k-}corrections in the V and K-band as
a function of age and $z$ for different values of 
$z_{f}$ in the range between 4 and 10 can be found at {\it http://www.brera.inaf.it/utenti/marcella/}.
As in the case of the \mastol\ ratios, at the above address
we also propose a set of  {\it reference} values of both the classic and the colour {\it k-}corrections
as a function of age [Gyr] and $z$.
In \S3.3 we supply with an easy use of all those  numbers, proposing the fit
of $k-$corrections in the K band as a function of age and of $z$ (assuming$z_{f}=4$).

\begin{figure}
\centering
\includegraphics[width=8.7truecm]{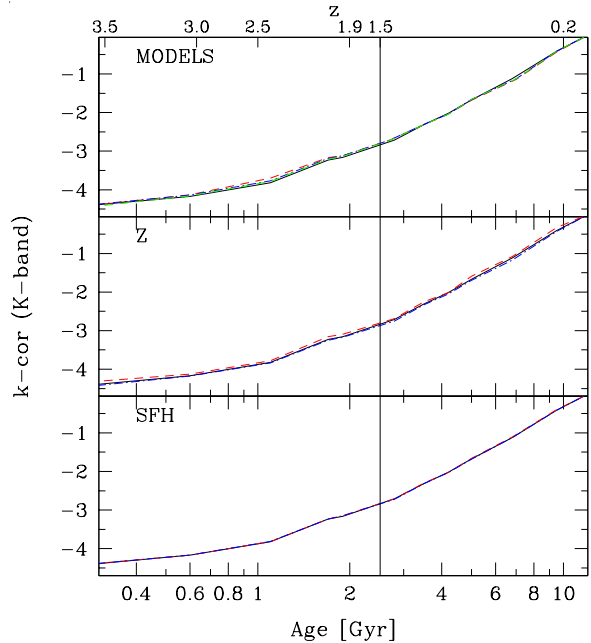}
 \caption{Colour {\it k-}correction in the K-band, obtained from the observed
nearest band, from K to Spitzer 8.0$\mu$m filters.}
\end{figure}

\begin{figure}
\centering
\includegraphics[width=8.7truecm]{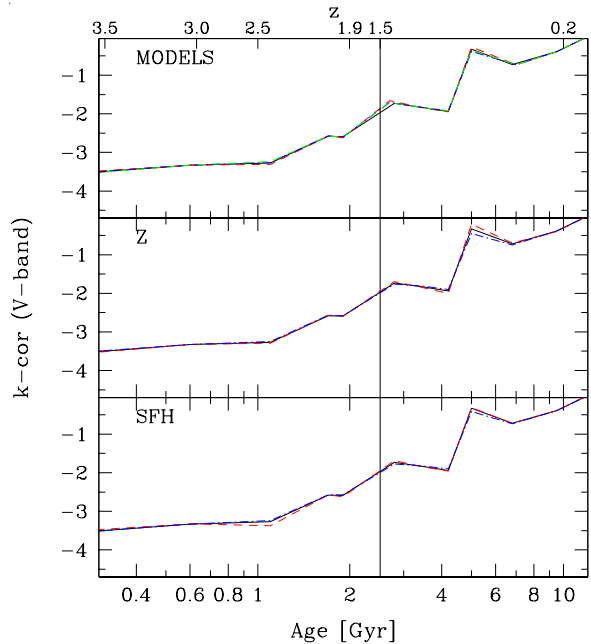}
 \caption{Colour {\it k-}correction in the V-band, obtained from the observed
nearest band, from V to K filters.}
\end{figure}

\subsubsection{Masses from apparent magnitudes: useful relations}
In the previous two subsections, we have analyzed the behavior of
the parameters \mastol\ and {\it k-}corrections as a function of models and model parameters,
suggesting the use of near-IR bands. We have also seen that IMFs, metallicities
and codes at a fixed age can produce values of one or both of the two quantities which
differ by a large amount with respect to the {\it reference} values. In this section,
we want to check the behavior of the combination of the two quantities included
in the formula that transforms apparent magnitudes into stellar masses (see equation 6):
$$\log(\mathcal{M/L})+0.4kcor$$
Figure 10 shows the behavior of such quantity as a function of age [Gyr] and
$z$ (assuming $z_{f}=4$), for different combinations of models and IMFs.
In particular, it is remarkable that the values of this quantity calculated
with Ma05 code and Salpeter IMF well agree with the values which can be calculated
with BC03 models. Indeed, the difference between the two codes as far as
the combination of {\it k-}corrections and \mastol\ in the K band are concerned
turns out to produce differences in the mass estimates at fixed apparent magnitude
which are of less than 20\% (between a factor 0.9-1.2) 
as a maximum and which are on average less than 5\%.
Something similar happens with the values calculated with CB08 code compared
with the values calculated with BC03 code and Chabrier IMF (see top panel of figure 10).
The differences between the two codes turn out to produce differences in the stellar
masses which are within a factor 0.8-1.2. At the same time, from the middle panel
it is evident that a larger discrepancy in the mass estimate can arise within models
built with the same code but with different metallicities. Adopting models
with $Z=0.2Z_{\odot}$ mass estimates (at fixed age) result on average less than 80\%
of the mass estimates derived assuming solar metallicity, while if $Z=2.5Z_{\odot}$
mass estimates result on average a factor 1.4 higher than the estimates at solar
metallicity.

\begin{figure}
\centering
\includegraphics[width=8.7truecm]{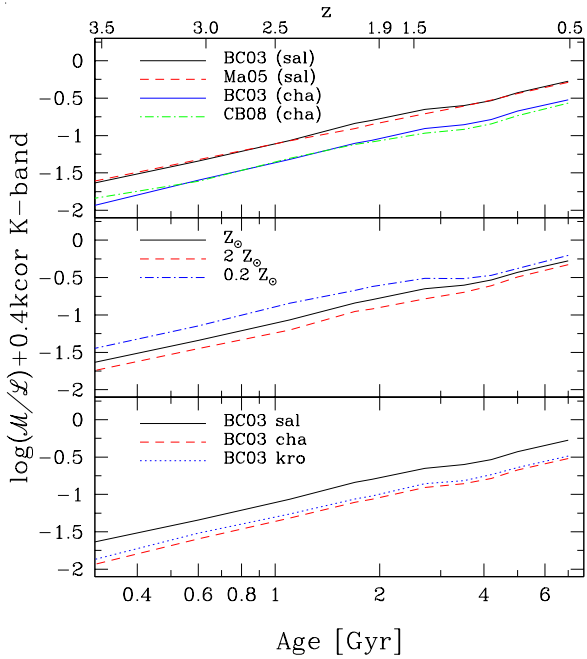}
\caption{ The quantity $\log(\mathcal{M/L})+0.4kcor$ is plotted as a function of age (bottom axis) 
and $z$ (assuming $z_{f}=4$, top axis). The upper panel compares values calculated with different models
at solar metallicity, the middle panel compares values obtained with BC03 models at different metallicities,
and the lower panel compares values obtained with BC03 model at solar metallicity but with different IMFs.
}
\end{figure}

\subsection{Stellar mass from different codes and parameters}
Details of the dependence of \mastol\ rations and {\it k-}corrections
on models and model parameters have been described in the previous subsections, and many
tables which summarize the corresponding values as a function of age and $z$ for
different values of $z_{f}$ in the range between 4 and 10
can be found at {\it http://www.brera.inaf.it/utenti/marcella/}.
Here we supply with an easy use of all those numbers, proposing the fits of some quantities
as a function of age and of $z$ for
$z_{f}=4$. 
In particular, Table 2 reports the values of the coefficients $a_{i}$ of the following 
expressions:

\begin{equation}
[\mathcal{M/L}_{K}]_{Y}=a_{0} + a_{1} \times  age + a_{2} \times age^{2} + a_{3} \times age^{3}
\end{equation}

\vskip 0.3truecm\noindent
$[\mathcal{M/L}_{K}]_{Y}=a_{0} + a_{1} \times z + a_{2} \times z^{2} + a_{3} \times z^{3} \hfill (9b)$

\vskip 0.1truecm\noindent
where Y stands for the following combinations of models and IMFs: 
{\it i)} Salpeter IMF and all the models excluded the CB08 and Ma05 ones, 
{\it ii)} Salpeter IMF and Ma05 model,
{\it iii)} Chabrier IMF and BC03 model, {\it iv)} Chabrier IMF and CB08 model, 
{\it v)} Kroupa IMF and BC03 model.
Note that the combination Salpeter IMF and Ma05 model, and Chabrier IMF and CB08 model have been
proposed separated from the other fits for the same IMF because they supply quite different values
of the \mastol\ ratio contrary to the homogeneity (within 20\%) of the results 
achieved with the other models. 
The fits minimize the percentage difference between the polynomial expressions of equations 9 and 9b
and the exact values, and they are all stopped to the minimum order that returns on average the exact values within 5\%.
As an example aimed at showing the accuracy of the fits mentioned above, one can consider that
a galaxy at $z=1.0$ with age=4.2 Gyr would result having $[\mathcal{M/L}_{K}]^{Sal}=$ 0.57, 
as average from BC03, PEGASE and GRASIL models. The values resulting from the fits are
0.57 and 0.58 as a function of age and of $z$, respectively. The same galaxy would have $[\mathcal{M/L}_{K}]^{Cha}=$ 0.35
and 0.36 and $[\mathcal{M/L}_{K}]^{Kro}=$ 0.37 and 0.39 (as a function of age and $z$, respectively)
to be compared with the values derived from BC03 models equal to 0.32 and 0.36.
The accuracy of the fit with respect of the \mastol$_{K}$ derived by the single models
can be seen in Figure 11 (top panel) for the case
{\it i)} as a function of $z$. Values derived 
with the Ma05 code are also reported for comparison, and it can be noticed that they are
higher at low redshift and lower at high redshift.

Table 3 reports the values of the coefficients $a_{i}$ of the following
expressions:

\begin{equation}
kcor_{Y}=a_{0} + a_{1} \times  age + a_{2}  \times age^{2} + a_{3} \times  age^{3} + a_{4} \times  age^{4}
\end{equation}

\vskip 0.3truecm\noindent
$kcor_{Y}=a_{0} + a_{1} \times  z + a_{2} \times  z^{2}+ a_{3} \times  z^{3} + a_{4} \times  z^{4} \hfill (10b)$

\vskip 0.3truecm\noindent
where Y stands for the following combinations of models and $Z$ values: 
{\it i)} all the models excluded CB08 and Ma05 at solar metallicity, which result
homogeneous within less than 10\%, {\it ii)} 
all the models excluded CB08 and Ma05 at $z=0.2 Z_{\odot}$, {\it iii)} all the models excluded CB08 
and Ma05 at $z=2.5 Z_{\odot}$, {\it iv)} Ma05 model at solar metallicity and {\it v)} CB08 model
at solar metallicity. Note that {\it k-}corrections depend on the metallicity of the stellar
populations but not on the IMF used to model them. Furthermore, Ma05 and CB08 models
supply quite different values of {\it k-}corrections in the K band, and thus separated fits have been
proposed for these models.

\begin{table*}
\caption{Coefficients of the fits of \mastol$_{K}$ as a function of age [Gyr] and $z$ ($z_{f}=4.0$):
$[\mathcal{M/L}_{K}]_{Y}=a_{0} + a_{1} \times $ age $+ a_{2}  \times $age$^{2} + a_{3}  \times $age$^{3}$ and
$[\mathcal{M/L}_{K}]_{Y}=a_{0} + a_{1}  \times z + a_{2}  \times z^{2} + a_{3}  \times z^{3} $
where Y stands for the following combinations: {\it i)} Salpeter IMF and all the models excluded CB08 and Ma05, 
{\it ii)} Salpeter IMF and Ma05 models,
{\it iii)} Chabrier IMF and BC03 models, {\it iv)} Chabrier IMF and CB08 models, {\it v)} Kroupa IMF and BC03 models.
Note that the combinations of Salpeter IMF with the Ma05 model, and Chabrier IMF with CB08 model have been
proposed separated from the other fits for the same IMFs because they supply quite different values
of the \mastol\ ratio contrary to the homogeneity of the results achieved with the other models.
}
\centerline{
\begin{tabular}{lrrrrccrrrrc}
\hline
\hline
 &     $a_{0}$ & $a_{1}$ & $a_{2}$ & $a_{3}$ & $\chi^{2}$ & \ \ &  $a_{0}$ & $a_{1}$ & $a_{2}$ & $a_{3}$ &  $\chi^{2}$ \\
\hline
 & \multicolumn{5}{c}{$[\mathcal{M/L}_{K}]_{Y}=a_{0} + a_{1} \times $ age $+ a_{2} \times $ age$^{2}+ a_{3} \times $ age$^{3}$ } & \ \ & \multicolumn{5}{c}{$[\mathcal{M/L}_{K}]_{Y}=a_{0} + a_{1}  \times z + a_{2}  \times z^{2} + a_{3}  \times z^{3}$ } \\
\hline
{\it i)}  & \multicolumn{11}{l}{Salpeter IMF} \\
 & 0.07  &  0.14  & -0.005 & 0.0000 &  0.01  & \ \ &  0.96 & -0.43 & 0.054  & 0.0000 &  0.03 \\
{\it ii)} & \multicolumn{11}{l}{Salpeter IMF - Ma05}                                              \\
 & 0.12 & 0.00 & 0.041 & -0.0032 & 0.07  & \ \ &   1.42 & -1.04  &  0.266 & -0.0211  &  0.02 \\
{\it iii)} & \multicolumn{11}{l}{Chabrier IMF - BC03}                                              \\
 & 0.03  & 0.10  & -0.008  &  0.0004 & 0.03    & \ \ & 0.67 & -0.47 & 0.146 & -0.0178 & 0.05 \\
{\it iv)} & \multicolumn{11}{l}{Chabrier IMF - CB08}                                              \\
 & 0.04  &  0.02 & -0.0104 & 0.00078  &  0.04    & \ \ & 0.64 & -0.522 & 0.1647 & -0.01814 & 0.04 \\
{\it v)} & \multicolumn{11}{l}{Kroupa IMF - BC03}                                                \\
 & 0.04  &  0.11  & -0.010  &  0.0005 & 0.04  & \ \ &  0.61  & -0.26  &  0.031  & 0.0000 & 0.07 \\
\hline
\hline
\end{tabular}
}
\end{table*}

Table 3 proposes also the fit of the colour {\it k-}corrections between observed 4.5$\mu$m
Spitzer band and rest frame K-band that stands for ages younger than 7 Gyr and/or $z>0.2$ 
(see details and definitions in \S3.2.2) which almost do not depend
on the choice of the models and on the metallicity of the stellar populations.
Following the same example as before, 
a galaxy at $z=1.0$ with age=4.2 Gyr, at solar metallicity would result having $kcor$= -0.77 as average value
of those derived from the BC03, PEGASE and GRASIL models. The values resulting from the fits are
-0.75 and -0.73 as a function of age and of $z$, respectively. 
The same galaxy at $Z=0.2 Z_{\odot}$ would have $kcor$= -0.93 in the BC03 models to be
compared with the values of -0.95 and -0.90 resulting from the fits as function of age  
and $z$, respectively. 
As in the case of \mastol$_{K}$, we report an example of comparison between data
and fit of eq (10b) in the middle panel of figure 11. Points derived from the 
Ma05 code are also reported and it can be noticed that at low ages (e.g. high redshift) they are much higher
than the average of the values obtained with the other models.

\begin{table*}
\caption{Coefficients of the fits of K-band {\it k-}corrections as a function of age [Gyr] and $z$ ($z_{f}=4.0$):
$kcor_{Y}=a_{0} + a_{1} \times $ age $+ a_{2}  \times $age$^{2} + a_{3} \times $ age$^{3} + a_{4} \times $ age$^{4}$ and
$kcor_{Y}=a_{0} + a_{1}  \times z + a_{2} \times  z^{2}+ a_{3}  \times z^{3} + a_{4}  \times z^{4}$
where Y stands for the following combinations: {\it i)} all the models excluded CB08 and Ma05 at solar metallicity, {\it ii)}
all the models excluded CB08 and Ma05 at $z=0.2 Z_{\odot}$, {\it iii)} all the models excluded CB08
and Ma05 at $z=2.5 Z_{\odot}$, {\it iv)} Ma05 model at solar metallicity and {\it v)} CB08 model
at solar metallicity. Note that Ma05 and CB08 models
supply quite different values of {\it k-}corrections in the K band, and thus separated fits have been
proposed. The last two lines list the coefficients of the same fits above but for the colour 
{\it k-}corrections between observed 4.5$\mu$m
Spitzer band and rest frame K-band that stand for ages younger than 7 Gyr and/or $z>0.2$
which almost do not depend
on the choice of the models and on the metallicity of the stellar populations.
}
\centerline{
\begin{tabular}{lrrrrrccrrrrrc}
\hline
\hline
 & $a_{0}$ & $a_{1}$ & $a_{2}$ & $a_{3}$ & $a_{4}$& $\chi^{2}$  & \ \ & $a_{0}$ & $a_{1}$ & $a_{2}$ & $a_{3}$ & $a_{4}$ &  $\chi^{2}$   \\
\hline
 & \multicolumn{6}{l}{$kcor_{Y}=a_{0} + a_{1} \times $ age $+ a_{2}  \times $age$^{2} + a_{3} \times $ age$^{3} + a_{4} \times $ age$^{4}$} & \ \  &
\multicolumn{6}{l}{$kcor_{Y}=a_{0} + a_{1}  \times z + a_{2}  \times z^{2}+ a_{3}  \times z^{3} + a_{4}  \times z^{4}$ } \\
\hline
 {\it i)} &\multicolumn{13}{l}{$Z=Z_{\odot}$} \\
 & -1.95 & 0.84 & -0.192 & 0.0143 & 0.0000 &  0.03 & \ \ & -0.32 & -0.65 & 0.314  & -0.0735 &  0.0000 &0.01  \\
 {\it ii)} &\multicolumn{13}{l}{$Z=0.2 Z_{\odot}$} \\
 & -2.17 &  0.77 & -0.159 & 0.0115 &  0.0000 &0.02 & \ \ & -0.35 & -0.81 & 0.309 & -0.0636 &  0.0000 &0.01 \\
 {\it iii)} &\multicolumn{13}{l}{$Z=2.5 Z_{\odot}$} \\
 & -1.76 &  1.67 & -0.703 &  0.1162 & -0.00655 & 0.01 & \ \ & 0.66 & -3.39 & 3.200 & -1.1595 & 0.13606 & 0.03 \\
 {\it iv)} &\multicolumn{13}{l}{Ma05 $Z=Z_{\odot}$} \\
 & -2.18 & 2.45 & -1.258 & 0.2429 & -0.01554 & 0.07 & \ \ & 0.20 & -2.46 & 1.828 & -0.4484 & 0.02245 & 0.01 \\
 {\it v)} &\multicolumn{13}{l}{CB08 $Z=Z_{\odot}$} \\
 & -2.10 & 2.50 & -1.1782 & 0.2124 & -0.01291 & 0.07 & \ \ & 0.98 & -4.90 &  4.805 & -1.7598 & 0.20933 & 0.07 \\
\hline
\hline 
 & \multicolumn{13}{l}{$m(4.5\mu)_{z}-m(K)_{z=0}=kcor$} \\
\hline
 & \multicolumn{6}{l}{$kcor_{Y}=a_{0} + a_{1} \times $ age $+ a_{2}  \times $age$^{2} + a_{3} \times $ age$^{3} + a_{4} \times $ age$^{4}$} & \ \ &
\multicolumn{6}{l}{$kcor_{Y}=a_{0} + a_{1}  \times z + a_{2}  \times z^{2}+ a_{3}  \times z^{3} + a_{4}  \times z^{4}$} \\
\hline
 & -3.32 &  0.33 & -0.004 &  0.0000 & 0.00000 & 0.01 & & -0.43 & -1.81 & 0.300 & 0.0000 &  0.00000 & 0.02  \\
\hline
\hline
\end{tabular}
}
\end{table*}

Finally, before concluding this section, we want to propose a very simple way to obtain
stellar mass estimates of early-type galaxies that combines the results retrieved
above on the \mastol\ ratio and {\it k-}corrections in the K band. 
Indeed, we have seen that for fixed IMF \mastol$_{K}$ and luminosities (by means of {\it k-}corrections)
only slightly depend on the specific code used to derived them, and their largest dependence is on age
and $z$ of the stellar populations. This means that for a given apparent luminosity,
stellar mass can be derived just as function of $z$ if we assume that age itself is a function of $z$
(e.g. $z_{f}=4$).  Thus, we propose the following equations:

\vskip 0.4 truecm
\centerline{--------------------------------------------------------------------------- }
\begin{equation}
log[\mathcal{M}^{Sal}(z)] = [-0.4\times m_{K}]+[17.45+1.22 z-0.25 z^{2}]
\end{equation}
\noindent
$log[\mathcal{M}^{Sal}(z)] = [-0.4\times m_{4.5}]+[17.52+0.49 z-0.10 z^{2}] \hfill (11b)$
\centerline{--------------------------------------------------------------------------- }
\vskip 0.1 truecm

\noindent
to derive $\mathcal{M}$ from the apparent K band magnitude $m_{K}$ (eq. 11) or 4.5$\mu$m band magnitude
(eq. 11b),   and from the redshift $z$,
assuming Salpeter IMF.
Equations (11) and (11b) return a mass estimate that results within 0.15 dex of
the value that could be obtained adopting all the models and the metallicities discussed in the previous section
(i.e. within a factor between 0.7-1.4),
including the Ma05 and CB08 models (see the discussion in \S3.2.3).
Indeed, Figure 11 (bottom panel) displays for all the models listed above (Salpeter IMF) the values
of  $\log(\mathcal{M/L}_{\lambda})+0.4kcor_{\lambda}+2\log(d_{pc})+-2.0+0.4 M_{\lambda}^{sun}$ 
as a function of the redshift $z$, which show a large homogeneity contrary to
the larger discrepancies found in the values of $kcor$ and of \mastol.

\begin{figure}
\centering
\includegraphics[width=8.7truecm]{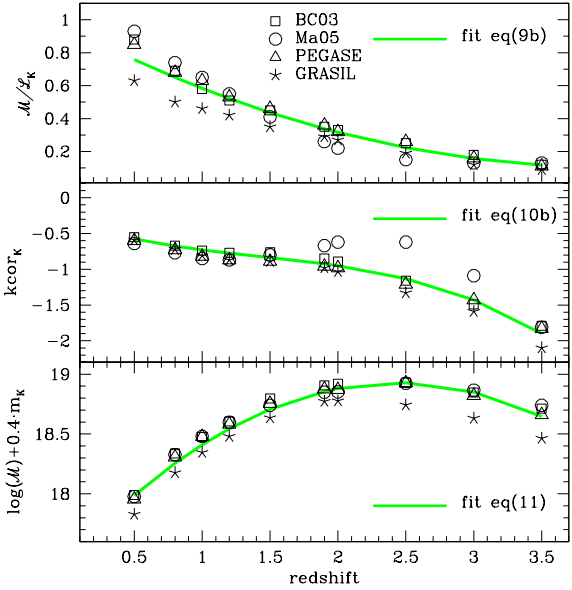}
\caption{{\it Top panel:} \mastol\ ratio in the K band for different models (Salpeter IMF)
as a function of $z$. The thick grey (green) line shows the fit of the data as in eq. (9b),
where the coefficients have been taken from Table 2 case {\it i)}. 
{\it Middle panel:} {\it k-}correction in the K band for different models (solar metallicity)
as a function of $z$. The thick grey (green) line shows the fit of the data as in eq. (10b),
where the coefficients have been taken from Table 3 case {\it i)}.
{\it Bottom panel:} the $z$-dependent part of the formula used to calculate stellar masses
(see eq. 6) from apparent K band magnitude $\log(\mathcal{M/L}_{\lambda})+0.4kcor_{\lambda}+2\log(d_{pc})+-2.0+0.4 M_{\lambda}^{sun}$
 is shown as a function of $z$, for different models
(Salpeter IMF and solar metallicity). The thick grey (green) line shows the fit of the data as 
in eq. (11).}
\end{figure}

\begin{figure}
\centering
\includegraphics[width=8.7truecm]{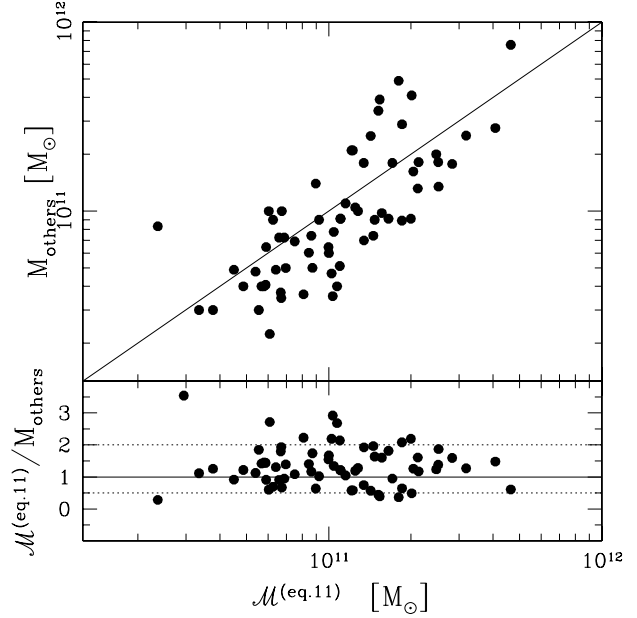}
\caption{ The stellar mass M$_{others}$ for a sample of 69 ETGs derived from the best 
fitting to their SEDs is compared with the mass $\mathcal{M}^{(eq.11)}$
derived by eq. (11). The 69 ETGs come from the samples of Rettura et al. (2006,
40 ETGs) and of Saracco et al. (2008, 29 ETGs). 
The mass $\mathcal{M}^{(eq.11)}$ has been scaled to the same IMF used by 
the authors using eqs. (12) and (13).
The solid lines represent the relation M$_{others}=\mathcal{M}^{(eq.11)}$.
For 55 out of the 69 ETGs the two estimates of the mass agree within a factor
2 (dotted lines, lower panel).} 
\end{figure}

\noindent
For instance, a galaxy at $z=1$ and age=4.2 Gyr with $m_{K}$=18.5 would result
having $1.0 \cdot 10^{11} \mathcal{M}_{\odot}$, to be compared with $1.2 \cdot 10^{11} \mathcal{M}_{\odot}$
for BC03 models, $1.2 \cdot 10^{11} \mathcal{M}_{\odot}$
for Ma05 models, $1.2 \cdot 10^{11} \mathcal{M}_{\odot}$
for PEGASE models and $0.9 \cdot 10^{11} \mathcal{M}_{\odot}$
for GRASIL models.
At higher redshift, a galaxy at $z=2$ and age=1.2 Gyr with $m_{K}$=19.0 would result
having $1.9 \cdot 10^{11} \mathcal{M}_{\odot}$, 
to be compared with $2.1 \cdot 10^{11} \mathcal{M}_{\odot}$
for BC03 models, $1.8 \cdot 10^{11} \mathcal{M}_{\odot}$
for Ma05 models, $1.9 \cdot 10^{11} \mathcal{M}_{\odot}$
for PEGASE models and $1.5 \cdot 10^{11} \mathcal{M}_{\odot}$
for GRASIL models. 

\noindent
It is worth noting that the above equations are based on the assumption of $z_{f}=4$.
On the other hand, results would continue to be in good agreement with those of equations
11 and 11b in case, for instance, of the assumption $z_{f}=6$. Indeed, the higher formation redshift
means that galaxies at the same $z$ are about 0.5 Gyr older, and the corresponding
{\it k-}corrections are less than 0.1 magnitudes higher. This difference in the
value of the {\it k-}corrections leads to a mass estimate that is 90\% of the value obtained
with $z_{f}=4$ as a maximum, that is well within the range 0.7-1.4 that is the uncertainty of equations 11 and 11b.

Masses obtained assuming Salpeter IMF can be simply converted into those with Chabrier
and Kroupa IMF adopting the following expressions:

\begin{equation}
[\mathcal{M}^{Cha}(z)] = 0.55 \times [\mathcal{M}^{Sal}(z)]
\end{equation}

\begin{equation}
[\mathcal{M}^{Kro}(z)] = 0.62 \times [\mathcal{M}^{Sal}(z)]
\end{equation}

\noindent
The conversion of $\mathcal{M}^{Sal}$ into $\mathcal{M}^{Cha}$ proposed above by means of
a constant independent of $z$ gives results which are within 10\% in agreement
with those obtained from BC03 models. $\mathcal{M}^{Cha}$ obtained by
means of the new CB08 models are lower than those obtained by means of BC03 models (between 70\% and 80\%)
and thus lower than the values resulting from equations (11) and (12) but still within
20\%. Finally, $\mathcal{M}^{Kro}$ resulting from equations (11) and (13) agrees within 10\%
with the values obtained by means of BC03 models with Kroupa IMF.

As a simple exercise, 
we compared the stellar mass derived from eq. (11) based on the apparent K-band
magnitude with the mass estimated from the SED fitting for a sample of 69 
ETGs at $0.5<z<2$ with spectroscopic confirmation of their redshift
and spectral type.
This sample includes 29 out of the 32 ETGs studied by 
Saracco et al. (2008)  and 40 ETGs taken from the 
compilation of Rettura et al. (2006) including some ETGs 
of the sample of van der Wel  et al. (2005) and of di Serego 
Alighieri et al. (2005).
The 29 ETGs taken from Saracco et al. (2008) have stellar masses obtained
with Chabrier IMF while those from Rettura et al. (2006) are based on 
Kroupa IMF.
To compare these masses with those derived from eq. (11) tuned on Salpeter IMF
we used eqs. (12) and (13) to scale them to the same IMF. 
The comparison is shown in Fig. 12 where the original mass  M$_{others}$  
provided by the other authors is plotted versus the mass 
$\mathcal{M}^{(eq. 11)}$ (upper panel).
In the lower panel the ratio between the two masses is shown.
The agreement between them  is remarkable: for 55 out of the 69 ETGs 
(80\% of the sample) the two masses agree  within a factor 2 while
they agree within a factor 3 for 66 ETGs (96\% of the sample).

\section{The best fitting technique applied to early-type galaxies at $1<z<2$}

The previous section has described the differences of the mass estimates obtained
with different codes or different parameters within the same code at fixed age and redshift.
But when multi-wavelength photometric data are available, age is not associated {\it a priori}
to the galaxies. On the contrary, models are used to build libraries of spectral templates 
which are adopted to find the one best fitting the whole SED of the galaxies. In the best fitting
technique, age is the fundamental parameter that is determined and that fixes the model
reproducing the observed SED of the galaxy. 
With the aim to assess the reliability of the estimates of the main physical properties 
of early-type galaxies at $1<z<2$ derived by the technique of best fitting 
a wide set of photometric data,
we apply it to a set of simulated galaxies for which the input properties are known.
We then compare the obtained results  with the known true values, focusing
our attention on the stellar mass estimate  and its dependence on the different mass estimators
used to derive it. We want to emphasize here that the following analysis
considers results obtained by means of photometric data only, while possible spectroscopic
information could in principle allow to obtain better results.

\begin{table}
\caption{Model parameters for the mock catalogs. The symbol X marks the
combinations among ages, redshift and times of the secondary burst which have been considered.}
\centerline{
\begin{tabular}{lccccc}
\hline
           & \multicolumn{5}{c}{ages [Gyr]} \\
\hline
$z=1.0$                   & 4.2 & 3.0 & 2.7 & 2.0 & 1.7 \\
no burst                  & X   &  X  &  X  &  X  &  X  \\
at $t=0.6$ Gyr            & X   &  X  &  X  &  X  &  X  \\
at $t=1.7$ Gyr            & X   &  X  &  X  &  X  &     \\
at $t=2.7$ Gyr            & X   &  X  &     &     &     \\
at $t=3.5$ Gyr            & X   &     &     &     &     \\
\hline
$z=1.5$                   & 4.2 & 3.0 & 2.7 & 2.0 & 1.7 \\
no burst                  &     &  X  &  X  &  X  &  X  \\
at $t=0.6$ Gyr            &     &  X  &  X  &  X  &  X  \\
at $t=1.7$ Gyr            &     &  X  &  X  &  X  &     \\
at $t=2.7$ Gyr            &     &  X  &     &     &     \\
at $t=3.5$ Gyr            &     &     &     &     &     \\
\hline
$z=2.0$                   & 4.2 & 3.0 & 2.7 & 2.0 & 1.7 \\
no burst                  &     &     &     &  X  &  X  \\
at $t=0.6$ Gyr            &     &     &     &  X  &  X  \\
at $t=1.7$ Gyr            &     &     &     &  X  &     \\
at $t=2.7$ Gyr            &     &     &     &     &     \\
at $t=3.5$ Gyr            &     &     &     &     &     \\
\hline
\hline
\end{tabular}
}
\end{table}

\vskip 0.3truecm
As a first step, we built three mock photometric catalogs of galaxies, whose properties 
in terms of wavelength coverage and photometric accuracy have been simulated according to those of two main
current surveys, the VIMOS VLT Deep Survey (VVDS, Le Fevre et al. 2005) and the
Great Observatories Origins Deep Survey (GOODS) for which we adopted the MUSIC sample (Grazian et al. 2006).
It is worthy to note that our aim is just to use the properties of these two surveys
to simulate the observations of a sample of early-type galaxies, and not to reproduce the observed
surveys themselves.
In particular, the wavelength coverage of two of our three mock catalogs corresponds to the
photometric bands in common to the two original catalogs, that are U, B, V, I, J and K, even
if filters slightly differ among the two surveys due to different instruments used for the observations
(e.g.  B, V and I band of the GOODS catalog are from HST-ACS F435W, F606W and F775W
filters, while the same bands in the VVDS are derived from observations with the Canada-France Telescope
and CFH12k mosaic camera). A third mock catalog adds to these six photometric bands the
four Spitzer-IRAC IR bands (3.6$\mu$m, 4.5$\mu$m, 5.8$\mu$m and 8.0$\mu$m) which are included
in the original MUSIC GOODS catalog.
As far as the photometric accuracy is concerned, we reproduced that of 
the two surveys in each band and in bins of 1 magnitude, for selected objects in the original 
catalogs with redshift $z>1$. Figure 13 shows the errors 
associated to each bin of magnitude in each of the 6 bands. The errorbars 
display the dispersion found around the average values of the errors,
and they depend on the number of objects populating each bin of magnitude in each of the two
catalogs. Since the photometric accuracy
only depends on the observed apparent magnitude and not on the spectral type of the
observed galaxies, we adopted the full VVDS and GOODS original catalogs to calculate
the expected errors. 

\noindent
The expected magnitudes in all the filters of the three
mock catalogs are calculated on a set of model templates whose parameters reproduce
the properties of early-type galaxies at $1<z<2$, assuming the
BC03 code at solar metallicity and with Salpeter IMF. The assumed SF history is described
by an exponentially declining SFR with time scale $\tau=0.6$ Gyr.
The possibility
of secondary star forming episodes is taken into account by the superimposition of $\tau=0.1$ Gyr
models on the principal SF history at different times assuming their contribution
to the final total stellar mass as equal to 5\% and to 20\%. 
The chosen combinations among ages, redshift and times of the secondary burst
(which are summarized in Table 4) constitute 57 template spectra,
among which 11 with no secondary
star forming episodes and 16 with recent (i.e.  $<1$ Gyr) starburst,
half of which forms 5\% of the total final mass while the contribution of the other half
is 20\%.

Finally, the stellar mass 
content of the simulated galaxies, needed to calculate their apparent magnitudes
as a function of their redshift, has been fixed as $0.3, 1.0, 3.0 \times10^{11}$ M$_\odot$,
and the possibility of dust obscuration has been taken into account assuming
A$_{V}=0.0,0.2,0.5$ and the reddening law of Calzetti et al. (2000).
In the catalogs we did not include the exact values of the calculated
apparent magnitudes, but values randomly generated following a gaussian distribution probability
around the exact value and with $\sigma$ equal to the assumed photometric errors.

The photometric catalogs of the simulated galaxies have been compared with those
of real samples of early-type galaxies, in order to test the reliability of the simulations.
The comparison sample has been collected considering the early-type galaxies
found in the GOODS catalog, following their spectral classification as reported by the 
last release of the spectroscopic catalog of GOODS (Vanzella et al. 2008)
 and/or by the K20 catalog (Mignoli et al. 2005).
Since most of the GOODS galaxies spectrally classified as early-types are at $z<1.2$,
we added some data of early-type galaxies at higher redshift selected from the
compilation of Saracco et al. (2008; see {\it http://www.brera.inaf.it/utenti/saracco/}), for a
total sample of 125 galaxies in the redshift range $0.6 < z < 2.2$.
Figure 14 (upper panels) shows the comparison of some colours of the simulated galaxies (reported as filled symbols) 
with those observed (crosses and open circles). In particular, the left panel displays the (V-K) vs. (V-I) colors,
while the right panel proposes the (V-K) vs. (J-K) colours. The arrows in the right-bottom
portion of each panel displays the direction along which a galaxy move in these diagrams
when the redshift, the age and the extinction are increased. As it can be seen, the simulated galaxies
well covers the parameter space of the observed galaxies.

\begin{figure}
\centering
\includegraphics[width=8.7truecm]{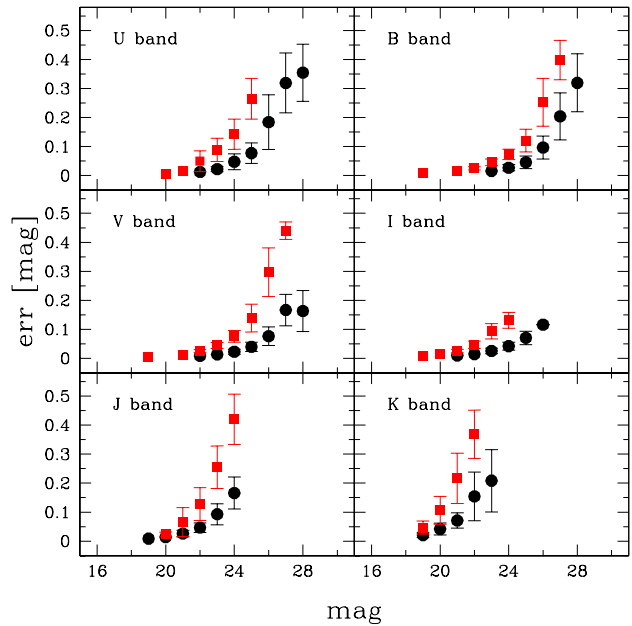}
\caption{Observed photometric errors associated to
the measured magnitudes in VVDS (grey/red filled squares)
and GOODS (black filled circles) original catalogs. The errorbars
represent the dispersion of the distribution of errors in each bin of
magnitudes.
}
\end{figure}

\begin{figure}
\centering
\includegraphics[width=8.7truecm]{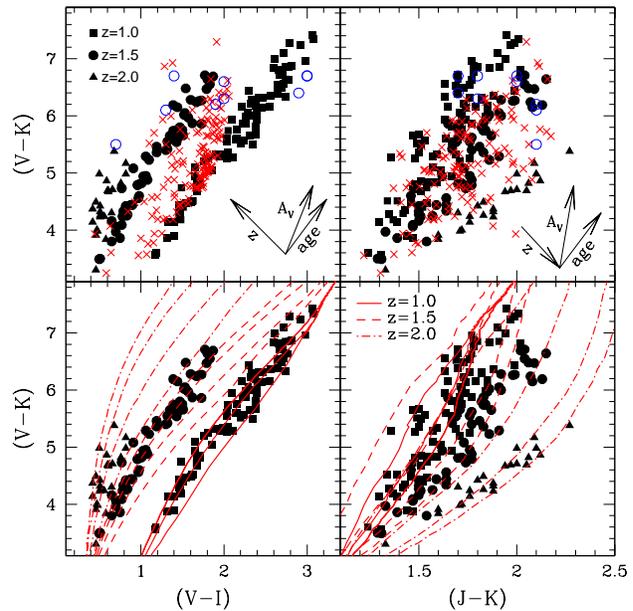}
\caption{Colour-colour diagrams of the simulated galaxies compared with a sample
of observed galaxies ({\it top panels}) and with models used to build the template spectral 
library adopted in the best fitting procedure with {\it hyperz} (see details in the text) 
({\it lower panels}).
Simulated galaxies are reported as filled symbols, which are squares for $z=1.0$,
circles for $z=1.5$ and triangles for $z=2.0$. Observations coming from the GOODS
sample are reported as crosses, while higher redshift data from Saracco et al. (2008)
are reported as open circles.  
Models sequences are reported as continuum lines (models at $z=1$), dashed line
(models at $z=1.5$), and dashed-dot lines (models at $z=2$).
The arrows in the right-bottom
portion of each panel display the direction along which a galaxy move in these diagrams
when the redshift, the age and the extinction are increased.
}
\end{figure}

\vskip 0.3 truecm
As a second step, we applied the best fitting technique to our three mock catalogs
by means of the photometric redshift code {\em hyperz} (Bolzonella et al. 2000). 
The code compares within a given range of redshift a set of spectral templates
with the observed photometric magnitudes and their uncertainties, varying
the free parameters: star formation time scale $\tau$, age, dust extinction. 
The spectral library adopted to find the best fitting template for each galaxy is composed by 
exponentially declining star formation models with time scales $\tau=0.1,0.3,0.6,1.0$ Gyr,
at solar metallicity. They have been generated by means of the BC03 code, assuming Salpeter IMF.
In the best fitting procedure the extinction
has been allowed to vary between A$_{V}=0.0$ and A$_{V}=0.5$
and at each $z$ ages have been
forced to be lower than the Hubble time at that $z$. Assuming that the redshift is known 
within 0.1, we run the code with $z$ varying between +/- 0.05 of its true value.
The assumed set of parameters reproduces the choices generally made to study the
photometric properties of real early-type galaxies with known spectroscopic redshift
(e.g.  Longhetti et al. 2005). It can be noted that the set of parameters chosen
to find the best fit of our simulated galaxies in some cases exactly covers the range
of parameters adopted to simulate the galaxies (e.g. A$_{V}$ values, IMF, metallicity), in other cases
it covers a smaller range (e.g. SF histories). Independently of the incompleteness
of the parameters adopted to build it, the template library allows to well cover the
colours of the simulated early-type galaxies. As an example, the lower panels of figure 14 show the comparison
of two colours sequences of the models adopted as template library in {\it hyperz} with the
colours of the simulated mock catalogs.

\vskip 0.3 truecm
Finally, once a best fitting template has been associated to each simulated galaxy, we have estimated
its luminosity, its \mastol\ ratio and its stellar mass content as derived
from the parameters of the best fitting template itself.
The obtained values  have been then compared with the true known input values of the same parameters
defining the model used to simulate the galaxies of the mock catalogs. In the following, we
analyze in details this comparison, for different bands and different estimators
both of luminosity and of stellar mass, while in Appendix A the same comparison
is analyzed for the \mastol\ parameter.

\subsection{Luminosity in the K and V bands} 

We derived {\bf $M_{K}$} and {\bf $M_{V}$} absolute magnitudes in the K and V bands
respectively, from the apparent 
$m_{V}$ and $m_{K}$ magnitudes, applying the {\it k-}corrections calculated on the best fitting templates, and
the dust corrections $dcor$ derived from the best fit values of A$_{V}$, besides the distance modulus:
$$M=m-d\_mod(z)-kcor(\lambda)-dcor(\lambda)$$ where $m$ is the apparent magnitude,
$d\_mod$ is the distance modulus ($d\_mod=5\log(d_{Mpc})+25$) 
and $kcor(\lambda)$ and $dcor(\lambda)$ are the {\it k-}correction and 
the extinction correction depending on the wavelength of the photometric band. 
We also calculated {\bf $M_{K_{4.5}}$} and {\bf $M_{V_{J}}$} which are K and V band absolute magnitudes 
derived from the apparent 4.5$\mu$m and J band fluxes respectively, i.e. from the observed bands closer
to their rest frame wavelengths. 
Finally, we define {\bf $M_{K_{raw}}$}=$m-d\_mod(z)-kcor_{raw}$ where $kcor_{raw}$
is the value calculated by means of equation (10b) and assuming the coefficients from Table 3
relative to the case {\it i)} valid for $Z=Z_{\odot}$ and all the models except Ma05 and Cb08 ones
(i.e. first line, right part of the table).

Figure 15 describes the uncertainties of the luminosity estimators introduced above.
The average value of the difference between the absolute magnitudes as recovered by
the estimators and their true values are reported as empty points, while
the errorbars mark the {\it maximum} differences between the same quantities. 
Square points refer to luminosity estimators 
which use the template best fitting the available photometric data to derive the corresponding
{\it k-}correction,
while the circle refers to $M_{K_{raw}}$ that does not need any multi-wavelength data fitting. 
For comparison, we report also
the range of uncertainties of the apparent magnitude used to derive the absolute one (shaded areas).
As far as the K band is concerned, K band magnitudes are generally recovered within 0.2-0.3 magnitudes
in the case of both the two  GOODS mock catalogs ($\Delta$K$<0.1$), 
while due to larger errors in the apparent magnitudes
($\Delta$K=0.2-0.3) the uncertainty is around 0.5 magnitudes in the case of the VVDS mock catalog, and it
can raise to more than 1.0 magnitude in the case of the fainter
objects (i.e.  with \mas$_{star}=0.3\times10^{11}$M$_{\odot}$ at $z=2$). Spitzer IR data with small
errors ($\Delta$(4.5$\mu$m)$<$0.1) allow to obtain estimates of the K band
magnitudes even better than 0.2 magnitudes.
It is particularly remarkable that the K$_{raw}$ value of the K band magnitude is well determined within 
0.3-0.4 magnitudes when the apparent K band  magnitudes are affected by small errors ($\Delta$K$<0.1$)
and within 1 magnitude for the larger errors of the fainter objects in the VVDS mock catalog.
This demonstrates the easiness to derive a reliable estimate of the near-IR luminosities
for early-type galaxies at known redshift.
Moreover, all the estimators of the K band magnitudes 
produce results which are 
distributed around the true 
values and systematic offset larger than few hundredths are not apparent.

\begin{figure}
\centering
\includegraphics[width=8.7truecm]{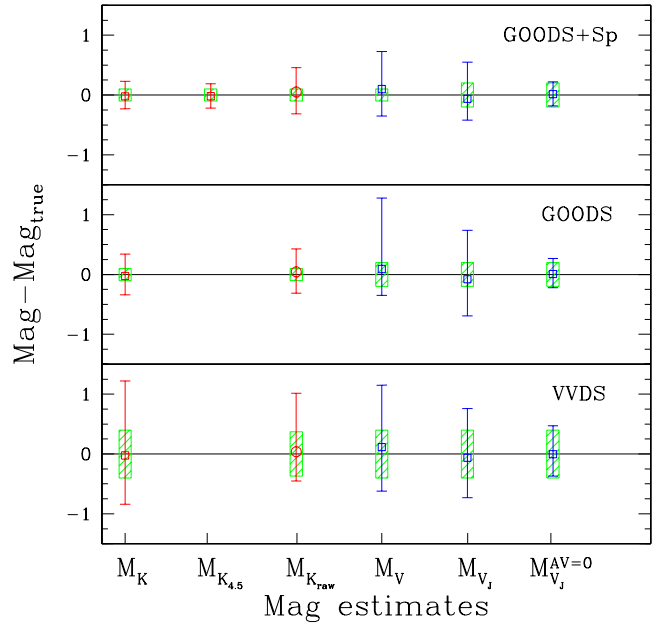}
\caption{Uncertainties in the estimates of the absolute magnitudes in the K and V band.
The reported errorbar represents the whole range of values spanned by the difference
between the true and the recovered magnitudes. Dots are the average values
of the same difference. Square dots refer to luminosity estimators
which use the best fitting template of the available photometric data to derive the corresponding
{\it k-}correction,
while the circle refers to $M_{K_{raw}}$ that uses a value of $kcor$ derived by equation (10b)
that does not need any multi-wavelength data fitting.
 The shaded areas highlight the range of uncertainties of the
apparent magnitudes used as the starting point to derive the absolute ones.
}
\end{figure}

\noindent
This is not the case in the V band, for which the value of $M_{V}$
statistically underestimates the true luminosities of the simulated galaxies of more than 0.1 magnitudes. 
Furthermore, the recovered values of $M_{V}$
show large differences with the true V absolute magnitudes, i.e.  larger than 1 magnitude.
The reason of the underestimate of the V luminosities is in the general underestimate of
the {\it k-}correction in this band (due to the trend of a slightly overestimate
of the age parameter), and at the same time its large uncertainty is 
due to the large uncertainty in the {\it k-}correction {\it and} dust
correction determination.
When $V_{J}$ is used to estimate the V luminosity, the uncertainty due to the
determination of the correct {\it k-}correction value disappears, and the
uncertainty due to the bad dust correction determination leads to slightly overestimate
on average the V band luminosity.
The last bar closest to the right border of figure 15 represents
results obtained assuming $A_{V}=0$ in {\em hyperz} for models
which do not include dust. $V_{J}^{A_{V}=0}$ retrieves the V absolute magnitude within 
0.3-0.4 magnitudes for the VVDS mock catalog, and 0.2 for the GOODS mock catalog,
that is within the uncertainties affecting the measured J band apparent magnitudes.
When real galaxies are observed, no information on the effective dust content
is known a priori, and thus the last results obtained fixing A$_{V}=0$ both
in models and in the templates used to analyze them is a simple exercise
to understand the behaviour of our optical magnitude estimates but it cannot
be considered representative of any serious data analysis.

Summarizing,
for the same goodness
of the best fit selected to represent the observations, the uncertainties in the determination 
of the V absolute magnitude are much larger than in the K band. 
In other words, it is easier to obtain a good determination
of near-IR absolute luminosities than optical ones.

\subsection{Stellar masses from different estimators}
Stellar mass content of the simulated galaxies has been recovered by means of several
different mass estimators:

\noindent
$\bullet$ 
$\mathcal{M}_{b}$ derives stellar masses from the scaling parameter $b$ of {\em hyperz}.
The parameter $b$ is the multiplying factor
needed to scale the model templates to match on average the observed available fluxes. In case of
the adopted code ({\em hyperz}) and of the assumed templates (the BC03 models), the stellar mass
content can be derived as: 
$$\mathcal{M}=b \times {2.0 \cdot 10^{-17}  4\pi d_{lum}^{2} \over{3.826\times 10^{33}}} \times \mathcal{M}_{mod}$$ 
where \mas$_{mod}$ is the mass corresponding to the best fitting template.

\noindent
$\bullet$ 
\mas$_{K}$ and \mas$_{V}$  are the stellar mass content derived from 
the mass to light ratios \mastol$_{K}$ and \mastol$_{V}$
corresponding to the best fitting template of each simulated galaxy,
and K and V absolute magnitudes $M_{K}$ and $M_{V}$ (defined as in the
previous subsection) respectively: 
$$\mathcal{M}_{i}=(\mathcal{M/L})_{i}\times 10^{-0.4(M_{i}-M_{i\ \odot})}$$ 
where $M_{i\ \odot}$ is the absolute magnitude of the
sun in the $i$-band (i.e.  $M_{K\odot}=3.41$ and $M_{V\odot}=4.83$, see Table 1).

\noindent
$\bullet$
\mas$_{K_{4.5}}$ and \mas$_{V_{J}}$ are derived from
\mastol$_{K}$ and \mastol$_{V}$ as in the previous case but use $M_{K_{4.5}}$ and $M_{V_{J}}$
(defined as in the previous subsection) 
as absolute magnitude in the K and V bands, respectively. 

\noindent
$\bullet$
In the V band, we also calculated
\mas$_{V_{J}}^{(B-V)_{0}}$ from $M_{V_{J}}$ and from the \mastol$_{V}$ ratio
derived from the rest frame (B-V)$_{0}$ colour following the Bell et al. (2005) prescription:
$$log[(\mathcal{M/L})_{V}]=-0.628+1.305 (B-V)_{0}$$
It is important to note that this relation has been introduced by the authors on
the basis of a calibration on samples of star forming galaxies,
while we have applied it to sample of simulated early-type galaxies.
The (B-V)$_{0}$ colour requested by the Bell et al. (2005) relation is the restframe
one, and thus it must be derived from the best fitting template of each
galaxy.

\noindent
$\bullet$
Finally, we estimated {\bf \mas$_{Kraw}$} that is the stellar mass content obtained
assuming equation (11).
This last estimate does not need any fitting since
it uses standard constant values depending only on the redshift. We want to check
the reliability of this mass estimator that can be useful when the early-type
nature of the galaxy is assessed and its redshift is known, but poor photometry
is available.

\begin{figure}
\centering
\includegraphics[width=8.7truecm]{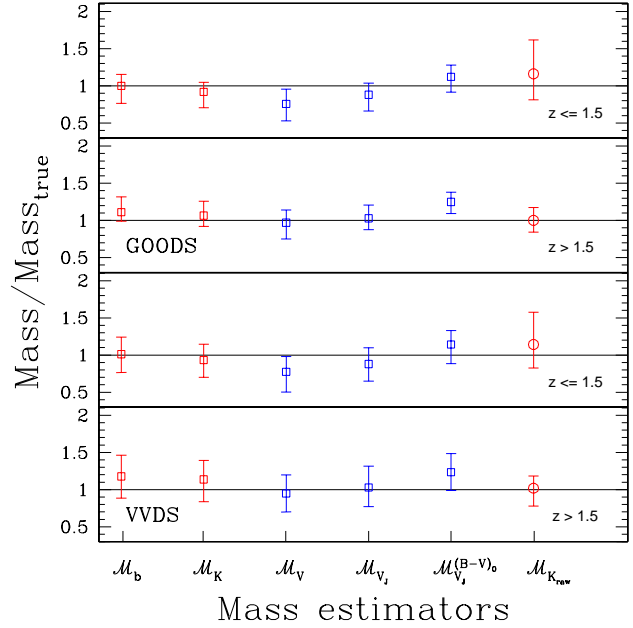}
\caption{Retrieved mass values obtained by means of different estimators
are reported normalized to their true values. 
Open  points represent the median values of the ratio between
the retrieved mass value and the true one, squares stand
for mass estimates depending on the best fitting template while
the circle stands for the only estimate that does not need any multi-wavelength data fitting 
(i.e. {\it raw} value, derived from eq. (9)). The errorbars represent the
16th (lower bar) and 84th (upper bar) percentiles of the distribution of
the same ratios, that
means that the 68\% of the results lie within the errorbars.
The upper two panels summarize the results obtained with the GOODS mock catalog,
while the lower two one those obtained with the VVDS mock catalog.
Results are presented for simulated galaxies at $z\le1.5$ (upper first
and third panels) and at $z>1.5$.
}
\end{figure}

\begin{figure}
\centering
\includegraphics[width=8.7truecm]{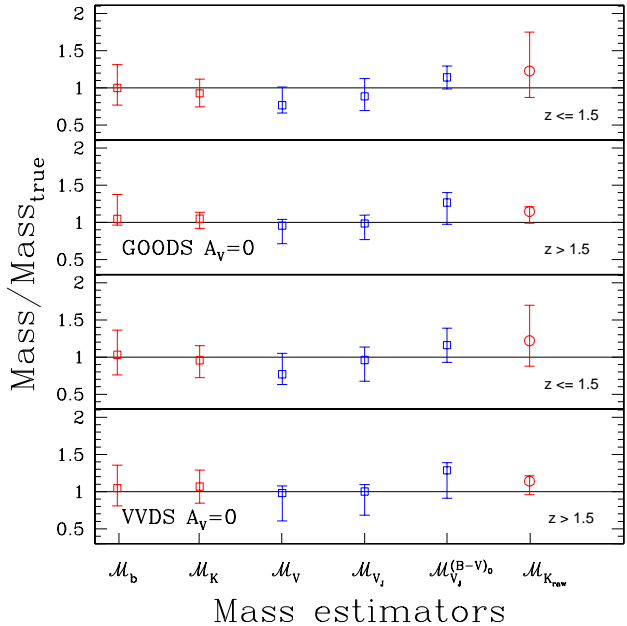}
\caption{The same as in figure 16, but considering only simulated
galaxies with A$_{V}=0$ analyzed assuming no dust extinction.
}
\end{figure}

\vskip 0.1truecm
\noindent
Figure 16 and 17 summarize the results obtained for the sample of simulated
galaxies when dust is a free unknown parameter and when extinction effects
are excluded, respectively.
Open  points represent the median value of the ratio between
the retrieved mass content and the true one, and they represent the trend
of each mass estimator to well reproduce the true stellar mass content
of galaxies or to over or under estimate it. The errorbars represent the
16th (lower bar) and 84th (upper bar) percentiles of the distribution of 
the ratios between retrieved and true values of stellar mass content, that
means that the 68\% of the values lie within the errorbars.
The comparison between the upper two panels with the lower two ones
allows to appreciate the influence of the photometric errors in the
mass estimates obtained with the reported estimators. The comparison between
results reported in the first and third upper panels ($z \le 1.5$) with those reported in the 
second and fourth lower panels ($z>1.5$) allows to understand the behaviour of each 
mass estimator when different ranges of age are considered, because at
$z>1.5$ the stars populating the galaxies are forced to be younger the 2 Gyr.
Finally, the comparison between the results reported in figure 16 with the
corresponding ones reported in figure 17 allows to understand the role 
of the lack of knowledge of the dust extinction affecting the luminosities
of the galaxies when deriving their stellar masses.

\subsubsection{$\mathcal{M}_{b}$}
The best mass estimates can be derived by \mas$_{b}$ that 
retrieves values between 0.7 and 1.2 (0.5 and 1.5) of the true ones 
for 68\% (100\%) of
the simulated galaxies for both the VVDS mock catalog and the GOODS one at
$z \le 1.5$.
Simulated galaxies in the VVDS mock catalog at $z>1.5$ (i.e., with
large photometric errors for the fainter galaxies) still obtain the
correct estimate within 0.8 and 1.5 (0.5 and 2.0) with this estimator.
Another advantage in using \mas$_{b}$  as mass
estimator is that the probability of an overestimate of the true mass
content of galaxies is the same of its underestimate. Indeed
the median values of the distribution of \mas$_{b}$/\mas$_{true}$ 
is 1 within 0.07, and only for galaxies with large photometric errors
it becomes larger than 1
(e.g. for the VVDS simulated galaxies at $z>1.5$  is 1.18,
while for the GOODS ones at $z>1.5$ is 1.11). 
Figure 17 reveals that the reason of this overestimate of the mass content
is related to the possible overestimate of the
dust extinction. Indeed, when this effect is removed, even the faintest
galaxies obtain on average the correct estimate of their mass content
by means of \mas$_{b}$.

\subsubsection{$\mathcal{M}_{K}$}
Another good mass estimator is \mas$_{K}$ that displays an uncertainty similar 
to that of \mas$_{b}$, that is it retrieves the correct value of the mass
content within 0.7 and 1.2 (0.5 and 1.5) for 68\% (100\%) of the simulated samples
at least at $z\le1.5$. For the faintest objects, like those in the VVDS mock
catalog at $z>1.5$, \mas$_{K}$ is even more precise than \mas$_{b}$ when considering
the whole range of results.
On the other hand, \mas$_{K}$ displays a trend to slightly underestimate the
true mass content of galaxies, and the median value of \mas$_{K}$/\mas$_{true}$
is 0.9 for $z\le1.5$ galaxies and it becomes 1.0 only for the youngest
ones ($z>1.5$) at least in the GOODS mock catalog where photometric errors
are smaller. The same effect is appreciable in figure 17 that means that
it is not due to wrong dust extinction estimate. The fact that the underestimate
of the mass content is found only for galaxies at $z\le1.5$ reveals that the
problem is related to the age estimate of the galaxies, that at lower redshift, where a larger
range of acceptable ages is allowed, is slightly underestimated on average.
Lower ages means lower \mastol\ ratios as well shown in Appendix A.
Indeed, we verified a median difference between the retrieved ages and the true ones
of 0.3 Gyr for $z\le1.5$ while it is 0.0 at higher redshift.

\subsubsection{$\mathcal{M}$ from V band estimators: \mas$_{V}$, \mas$_{V_{J}}$ and \mas$_{V}^{(B-V)_{0}}$}
A general trend to underestimate the mass content of galaxies is displayed
by the mass estimator \mas$_{V}$, that retrieves on average a value of mass that is
only 75\% of the true one. The underestimate is less evident at
$z>1.5$ (where ages are better estimated) but it is still appreciable (\mas$_{V}$ is
90\% of the true mass content).
The reason of this trend is not related only to the possible wrong dust extinction
estimate because it is confirmed even in figure 17 where dust effects are 
excluded. Indeed, a large part of this effect is caused by the underestimate of the V band luminosity
by means of M$_{V}$ (see figure 15). The luminosity underestimate, that is of 0.1 mag on average,
leads to underestimate the mass content to about 0.9 of
its true value, that matches the underestimate found at $z>1.5$. When $z\le1.5$
we have already mentioned for the K band that the underestimate of the age of the stellar
populations produces the underestimate of the \mastol\ ratio. In the V band the effect is
even larger (see figure A1) and this explains the strong underestimate found
at $z\le1.5$ in the mass value derived by means of \mas$_{V}$.
The mass estimates obtained by means of \mastol$_{V}$ but which adopt the V band magnitude derived by the
observed J band, \mas$_{V_{J}}$, do not suffer of the same problem of underestimate of \mas$_{V}$, and they
are closer to the correct values. At $z>1.5$ \mas$_{V_{J}}$ retrieves the correct value of the
mass content of 68\% (100\%) of the galaxies within 0.8-1.2 (0.5-1.5) for the GOODS mock catalog and
within 0.7-1.3 (0.4-1.6) for the larger errors of the VVDS mock catalog.
At $z \le 1.5$ \mastol$_{V}$ still underestimates the true \mastol\ ratio in the V band because of the
underestimates of the ages, and \mas$_{V_{J}}$ is on average 85\% of the true value.
The use of \mastol$_{V}^{(B-V)_{0}}$ instead of \mastol$_{V}$ makes the 
mass obtained by means of \mas$_{V}^{(B-V)_{0}}$ overestimated with respect
to the true value, being around 1.3 for galaxies at $z\le1.5$ and 1.2
for galaxies at higher redshift. On the other hand, the 16th and 84th percentiles
for this mass estimator are close to the median value at least as those of the the previous ones.
Since Bell et al. (2005) have calibrated their \mastol\ estimator on a
sample of star forming galaxies, we believe that the same calibration is not valid
for quiescent galaxies. Indeed, our results demonstrate that the possible powerful
mass estimator \mastol$^{(B-V)_{0}}_{V}$ should be used only in the studies of        
young/star forming galaxies at low redshift, while its use in different context
could bring large errors and a general trend to strongly overestimate the stellar mass
of galaxies.

\subsubsection{\mas$_{Kraw}$}
The last mass estimator presented in figures 16 and 17 has the advantage that
it can be used without the need of any fitting, that is without a good photometry
needed to apply the best fitting technique. At $z\le1.5$ \mas$_{Kraw}$ retrieves masses
which are around 1.25 of the true values, and 68\%\ (100\%) of the galaxies obtain
stellar mass estimates within 0.9-1.8 (0.5-2.4) for both the two mock catalogs,
a much worse result than that obtained with the other estimators. On the contrary,
at $z>1.5$ \mas$_{Kraw}$ supplies values which have more or less the same confidence
of the best mass estimators such as \mas$_{b}$. Indeed, \mas$_{Kraw}$ at $z>1.5$
retrieves on average the correct value of the mass content of galaxies, and 68\%\ (100\%) of them
obtain estimates within 0.8-1.2 (0.6-1.6).
The reason of the bad results achieved at lower redshift is that at $z\le1.5$ the fit expressed in equation (11)
on which the estimator \mas$_{Kraw}$ is based takes into account only old ages (i.e. ages older than 2.7 Gyr),
while the simulated galaxies include also young (e.g. 1.7 Gyr old) galaxies which have a lower
\mastol\ ratio.

\vskip 0.4truecm
Summarizing, when photometric data are used to find the best fitting templates reproducing the observed
early-type galaxies, the best mass estimate, not affected by any systematic trend, is
that derived by the scaling factor between templates and data on average. Alternatively,
the near-IR bands can be safely used to derive the mass content of early-type galaxies, while
optical bands produces much larger uncertainties and general underestimate of the stellar
masses which are difficult to be taken into account.

\begin{figure}
\centering
\includegraphics[width=8.7truecm]{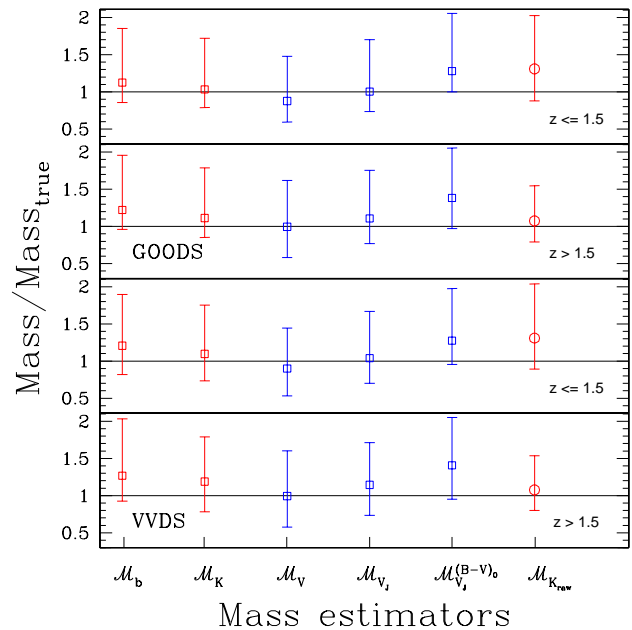}
\caption{The same as in figure 16, but results are derived from mock catalogs 
built on the basis of an enlarged set of metallicities and of model codes (i.e. 
PEGASE, MA05 and GRASIL, see details in \S4.2.4).
}
\end{figure}

It is worth to note that all the comparisons made in this subsection are
based on a fixed and known IMF (i.e. Salpeter IMF), and thus the uncertainties
reported do not include the large uncertainty due to the lack of knowledge
of the real IMF that produces real SEDs. At the same time, as we have seen
in \S3.2, IMFs can in many cases be considered scaling factors of the total stellar content
at fixed luminosities.
Anyhow, reported uncertainties for the mass estimates and for the luminosities are
surely underestimated with respect to real data analysis also for another reason.
Indeed, in the above test observations are simulated with templates that are contained
in the spectral set used to analyze data, while in the reality
synthetic templates are only approximation of the properties of real galaxies.

In order to take into account this point,
we built an other set of simulated galaxies on the basis of templates including
different metallicities and of the other codes listed in \S2.2 (i.e.,
PEGASE, Ma05, GRASIL). All the galaxies are modeled assuming Salpeter IMF and following
the combinations of ages and redshift already described at the beginning of this section.
We used this new enlarged set of early-type galaxies models at $1<z<2$
to build  new mock photometric catalogs, assuming again the uncertainties of the GOODS
and VVDS surveys.
The analysis of these new mock catalogs has been performed in the same way as the previously
built ones. In particular, we continued to adopt a set of templates
at solar metallicity and based on the BC03 code for the comparison in {\it hyperz}.
The results are presented in figure 18, and they can be directly
compared with those in figure 16.  As anticipated, errorbars (16th and 84th percentiles
of the distribution of the ratio \mas$_{retrieved}$/\mas$_{true}$) are larger in 
figure 18 than in figure 16, and their width is
almost the same (i.e. $\sim$0.7-0.8) for all the mass estimators presented. From
figure 18 it can be deduced that {\bf even at fixed and known IMF, the mass content
of the early-type galaxies at $1.0<z<2.0$ can be retrieved} within 0.7-1.5 (0.4-3.0) of
the true value
for 68\% (100\%) of the simulated sample, that is {\bf not better than a factor 2-3}.

\subsection{Different codes, different results?}
In this section we quantify the differences in the stellar mass estimates
obtained by means of the best fitting of the spectrophotometric data of 10 real
 early-type galaxies at $1<z<2$.
The mass estimators described in the previous section are used to calculate
the stellar content of 10 massive early-type galaxies at $z\simeq1.5$
using {\it hyperz} to fit their spectrophotometric data.

The sample of 10 massive galaxies is described in a series of papers (i.e.,
Saracco et al. 2003; Saracco et al. 2005; Longhetti et al. 2005).
The optical  (B, V, R, I) and near IR (J and K') magnitudes of the galaxies in the sample
are from the Munich Near-IR Cluster Survey  (MUNICS; Drory et al. 2001).
In the best fitting procedure we also include further 5 points derived from the observed spectra,
four in the wavelength range $0.9<\lambda<1.2$ $\mu$m and one in the H-band
(see details in Saracco et al. 2005). The best fitting procedure follows the original
one described in Saracco et al. (2005) and Longhetti et al. (2005), but four different
set of templates are here adopted. We refer to the ``BC03 set'' as the one based
on the BC03 code, including four SF histories
($\tau=0.1, 0.4, 0.6, 1.0$ Gyrs) and four values of the stellar metallicity
 ($Z_\odot$, 0.2$Z_\odot$, 0.4$Z_\odot$, and 2.5$Z_\odot$). The ``Ma05 set'' is
composed by templates built with the Ma05 code with fixed solar metallicity
and with three SF histories ($\tau=0.1, 0.4, 1.0$ Gyrs). The ``PEGASE set''
includes the same four SF histories as the BC03 one, but the metallicity within
each model is not fixed
and it depends on the age of the template selected. Finally, the ``Grasil set''
is composed by only one SF history, described in \S2.2, with stellar metallicity
that increases with time following a chemical evolution prescription as in the
case of the PEGASE set. All the four sets of models have been  built with
the Salpeter IMF to make easy comparisons among the different results obtained with
the different codes.

If we consider the best fit for each galaxy, mass estimates obtained 
with the Ma05 set and with each of the estimators previously described 
are on average $0.8\pm0.2$ with respect to the corresponding ones found with the BC03 set.
Recent works on mass estimates which compare results based on different set of templates
claimed that masses derived with the Ma05 set can be between 0.4 and 0.6 of the 
masses derived with the BC03 set (Cimatti et al. 2008).
Indeed, the two apparently different findings are
not surprising, since the above analysis is based on a sample of galaxies
at $z\simeq 1.4$, which are on average older (i.e. ages $>1-2$ Gyr) than those
contained in the comparison sample of Cimatti et al. (2008, $z>1.4$).
Indeed, we checked (on the basis of simulated and not real galaxies) that at
ages younger than few Gyr, mass estimates with
the Ma05 templates are on average 0.6 of those derived with
the BC03 ones.

The PEGASE set of templates supplies reasonable fits of the data
only when dust is used as a varying parameter.
 Mass estimates obtained  with PEGASE templates are on average
larger (i.e. 1.1 and 1.3 with or without
dust extinction respectively) than those obtained with
the BC03 set of templates.

Finally, it is interesting to evaluate the results in the mass estimates that are
achieved using the GRASIL code to build the set of templates used in the
best fitting procedure. As already emphasized, the GRASIL code generates spectra
of early-type galaxies following a self consistent picture of galaxy evolution,
from the points of view both of star formation and of metal enrichment. 
Results are on average in very good agreement with those found with the BC03 set.

 \vskip 0.2truecm
Summarizing, the use of different codes, at least those listed and discussed here,
does not bring any significant change in the final results when stellar mass
content of early-type galaxies is calculated by means of best fitting the
spectrophotometric data. In particular, when masses are estimated by means of
comparison with the BC03 set of templates, the obtained values are on average between 0.8-1.3
of those which could be obtained with the other sets here considered.
On the other hand, the intrinsic uncertainty in the stellar mass estimate
is larger than the former range, being between 0.5-1.5 even in the case of
good quality data as those of GOODS (see \S 4.2),
and further differences can be found when different
mass estimators are adopted, as already outlined in the previous section.

\section{A new empirical mass estimator}
As seen in the previous section, stellar mass content of high redshift early-type galaxies
cannot be estimated better than a factor between 2 and 3 on the basis of photometric data depending on the quality
of the available photometric data and/or on the distance (i.e. apparent luminosity) of the
galaxies. All the estimators presented above require the choice of a template
reproducing the observed SED of the galaxies, on the basis of which
the \mastol\ ratio and/or the {\it k-}corrections needed to transform observed
fluxes into stellar mass estimates are derived.
In the following we present a new stellar mass estimator based uniquely on the apparent
magnitudes in the K and V bands, and thus that it does not require any SED fitting. 
As for the other estimators, a requested {\it a priori} condition is
that the redshift range of the galaxies is $1<z<2$ even if the
correct redshift value can be poorly known (e.g. photometric redshift).
As we will see, its precision is not lower than that of the other {\it classic} estimators,
and in case of poor photometry it is more stable resulting even more reliable than the previous ones.
On the other hand, with respect to the other {\it classic} estimators it presents
the not negligible advantage that it does not require the fitting of large
set of multi-wavelength photometric data.

\begin{figure}
\centering
\includegraphics[width=8.7truecm]{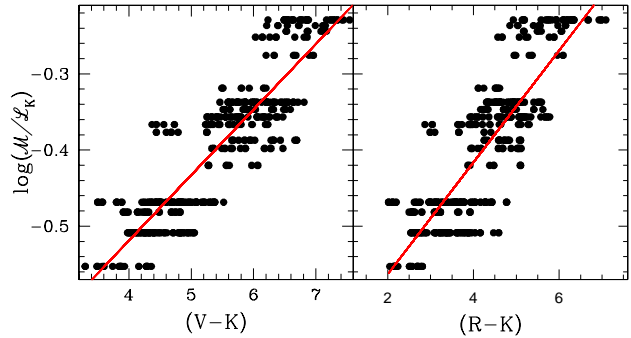}
\caption{The \mastol$_{K}$ ratio of the simulated galaxies is plotted
versus the (V-K) apparent colour ({\it left panel}) and the (R-K) apparent colour
({\it right panel}). The grey (red) line displays the fit of the data in both the two panels.
}
\end{figure}

\noindent
The new estimator is defined by the following equations:
\begin{equation}
\log[\mathcal{M}^{(V-K)}]=\log(\mathcal{M/L}_{K})-0.4\cdot M_{K} + 0.4\cdot M_{K}^{sun}
\end{equation}

\vskip 0.2 truecm
\noindent
where  

\vskip 0.2 truecm
\noindent
$0.4\cdot M_{K}^{sun}=1.364$ 
\vskip 0.2 truecm

\noindent
and

\vskip 0.2 truecm
\begin{equation}
\log[\mathcal{M/L}_{K}^{(V-K)}]=0.086\times(V-K)_{obs}-0.863
\end{equation}
\vskip 0.2 truecm

\noindent
while (V-K)$_{obs}$ is the observed (V-K) colour and $M_{K}$ is derived in a {\it raw} way, 
without any dust correction and assuming as {\it k-}correction value
the one calculated by means of equation (10b) assuming the coefficients in
Table 3 of case {\it i)}:

\vskip 0.2 truecm
\noindent
$M_{K}=m_{K}-d\_mod(z)-kcor(z)$

\vskip 0.2 truecm
\noindent
and

\vskip 0.2 truecm
\noindent
$kcor(z)=-0.32-0.65\cdot z-0.314\cdot z^{2}-0.0735\cdot z^{3}$

\begin{figure}
\centering
\includegraphics[width=8.7truecm]{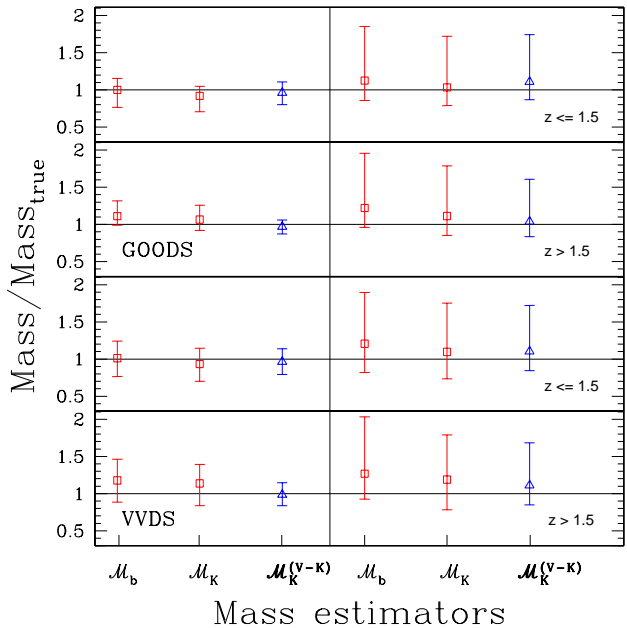}
\caption{Results obtained with the new near-IR mass estimator \mas$_{K}^{(V-K)}$,
defined on the basis of the K band luminosity derived in a {\it raw} way
(i.e., without any dust correction and assuming as {\it k-}correction value
the one calculated by means of equation (10b) assuming the coefficients in
Table 3 of case {\it i)}) and on the \mastol\ ratio that is a simple function of
the apparent (V-K) colour: $\log(\mathcal{M/L}_{K})=0.086\times(V-K)_{obs}-0.863$.
For comparison, we show also results obtained by means of \mas$_{K}$ and \mas$_{b}$
which contrary to \mas$_{K}^{(V-K)}$ require the fitting of the available multi-wavelength
photometric data. Symbols and errorbars have the same meaning of those in figure 16.
On the left, bars refer to results obtained on the mock catalogs built adopting
solar metallicity and BC03 code, while on the right bars refer to
the enlarged mock samples of galaxies simulated with different metallicities and
different model codes (see text).
}
\end{figure}

\vskip 0.2 truecm
\noindent
It is worthy to note that the equation defining the \mastol\ ratio in the K band
as a function of the (V-K) colour 
is totally empirical and it has been deduced
by the observation that the apparent (V-K) colour follows a quite tight relation
with \mastol$_{K}$ for the model parameters representing the class
of early-type galaxies (see \S2). Figure 19 shows the relation between
\mastol$_{K}$ and the (V-K) colour (left panel). Points represent
the simulated galaxies of the first set.
The scatter of the relation expressed in eq. (15) is $\sigma=0.04$ and the maximum
difference between the fit and the data of the simulated galaxies is $\pm 0.1$.

Figure 20 presents the results obtained with the new near-IR mass estimator \mas$_{K}^{(V-K)}$ (eq. 14 and 15),
for the mock catalogs
derived by the BC03 models at solar metallicity (third errorbars, left panels)
and for the enlarged mock catalogs derived by means of a wide range of metallicities
and of the other models listed in \S2.2 (third errorbar, right panels).
For comparison, we also show results obtained by means of \mas$_{K}$ and \mas$_{b}$
which contrary to \mas$_{K}^{(V-K)}$ require the fitting of the multi-wavelength
photometric data.
In spite of its simplicity, \mas$_{K}^{(V-K)}$ allows to obtain mass
estimates even more precise than \mas$_{b}$
 Indeed, the uncertainty introduced in the estimate of M$_{Kraw}$
(see Fig. 15) combined with that of this new \mastol\ estimator that
retrieves \mastol$_{K}$ within 0.9 and 1.1 (0.7 and 1.3) 
of the true values for 68\%\ (100\%) of the simulated sample, allows to
obtain mass estimates between 0.8-1.1 for 68\% of the
simulated galaxies, to be compared with 0.7-1.4 of the \mas$_{b}$ estimator.
Furthermore, \mas$_{K}^{(V-K)}$ is only slightly dependent on the photometric quality
of the data. Indeed, results are similar for the VVDS and GOODS data set, while
\mas$_{K}$ and \mas$_{b}$ show larger errors for the former sample with respect to the
latter one. Finally, as it can be deduced from the last three errorbars in each panel
of figure 19, \mas$_{K}^{(V-K)}$ retrieves the correct value of the stellar mass
content within 0.8-1.7 for 68\% of the galaxies even when simulated with a larger
range of metallicities and of synthetic codes, to be compared with 0.8-2.0 of \mas$_{b}$.
Thus, recalling again that the calculation of \mas$^{(V-K)}$ does not
require any fitting of large set of photometric data, we consider this mass
estimator as the one to be recommended, in particular when poor photometry is available.

As an exercise, 
we applied this new estimator to the sample of early-type galaxies at $z>1$ presented
by Saracco et al. (2008). We considered only galaxies for which V and K band magnitudes
were available, for a total number of 15 galaxies at $1.2<z<1.9$ and apparent K band magnitudes between
16.6 and 19.8. 
Table 2 in Saracco et al. (2008) 
reports also their redshift and their stellar mass content
derived by means of the $b$ best fitting parameter assuming the Chabrier IMF. 
We calculated for all these galaxies the stellar mass content by means of
\mas$_{K}^{(V-K)}$ as defined above, and we further applied a multiplying factor of 0.55
to transform Salpeter based mass estimates into Chabrier based ones (see eq. 12). For 12 out of the
15 galaxies (80\%\ of the sample), \mas$_{K}^{(V-K)}$ agrees with the values listed
in Table 2 within a factor of 2, that is well within the intrinsic uncertainty of the mass
estimate itself. For the remaining three galaxies the difference between the two estimates
is larger (e.g. a factor 3) and it reaches a factor 5 in one case. Since we do not know which is 
the true value of the stellar mass of the analyzed galaxies, we can conclude that the 
classic estimators and our new estimator provide discrepant
values  in less than 15\% of the cases.

Given its simplicity  and its stability with respect
to photometric inaccuracy/uncertainties, the mass estimator \mas$_{K}^{(V-K)}$ 
provides reliable and homogeneous mass estimates 
which can be easily compared among them since 
the dependence on models and on model parameters is cancelled.
For these reasons, we believe that it should be adopted for all the 
samples of {\it bona fide} elliptical galaxies at $1<z<2$ which are the
morphological type and the range of redshift for which the estimator has been tested. Obviously,
the larger is the error in the redshift estimate the larger is the corresponding error
in the mass estimate obtained with \mas$_{K}^{(V-K)}$, but the same is true of 
all the other estimators. An immediate application could be to the samples of EROs
(e.g. (R-K)$>$5) for which a morphological information is available (e.g. HST images)
and for which redshift can be deduced both on the basis of photometry and on the basis
of spectral information.

\noindent
Since the study of early-type galaxies is more often based on the (R-K) colour than
on the (V-K) one, we tried to test the validity of a similar relation to
obtain the K band mass to light ratio from (R-K). In the case of early-type galaxies simulated
only with the BC03 models at solar metallicity, a linear relation between \mastol$_{K}$
and the apparent (R-K) colour as 
\begin{equation}
\log(\mathcal{M/L}_{K}^{(R-K)})=0.073\times(R-K)_{obs}-0.710
\end{equation}
is not worse than the previous one between \mastol$_{K}$ and the apparent (V-K)
colour (see right panel of figure 19). 
Indeed the scatter of the relation expressed in eq. (16) is $\sigma=0.05$ and
the maximum difference between the fit and the data is $\pm 0.13$.
The problem arises when we consider galaxies modelled with other templates and
with other metallicities. While the relation involving (V-K) remains stable and it continues
to give good results, the one based on the (R-K) colour displays a much larger uncertainty.
In other words, the apparent (R-K) colour of simulated early-type galaxies at $1<z<2$ is sensitive to
the specific code and/or metallicity adopted to model the galaxies, while the (V-K)
colour remains more constant over model details.

\section{Summary and conclusions}
The aim of the present paper is to quantify the dependence of the estimates
of some basic properties of early-type galaxies, such as their luminosities and
stellar masses, on the different models and model parameters which can be assumed
to analyze the observational data. In other words, we want to answer to the question:
{\it do different model codes (and different model parameters assumed within the same code) 
produce significantly different results when they are used to derive
luminosity and mass estimates of early-type galaxies?}

\vskip 0.2truecm
In the first part of the paper, we analyzed the dependence of the \mastol\ ratio
and of the {\it k-}corrections for fixed age and {\it z} values from different model parameters
(such as metallicity and IMF) and different model codes among some of the most popular ones:
BC03, CB08, Ma05, PEGASE, GRASIL. We found that the best accuracy in the determination
of both the two quantities is achieved in the near-IR bands. 
Furthermore, even if CB08 and Ma05 models
give results different from the other codes when deriving \mastol$_{K}$ and K band {\it k-}corrections,
the combination of the same two quantities required to derive the stellar mass starting 
from the apparent K band magnitude is such that it results similar to that
calculated with the other codes. In other words, the stellar mass estimate derived
from the apparent K band magnitude assuming fixed IMF, at known age and {\it z}, is the same for
all the models here considered within 20\%. Equation (11) in \S3.3 summerizes
this finding by supplying the fit of $\log[$\mas$^{Sal}(z)]$ as function of $m_{K}$ and $z$.
This fit allows to recover within 20\% the stellar mass estimate that would have been
recovered adopting all the models here considered for the same value of $z$ and $m_{K}$.
Stellar masses derived assuming the Salpeter IMF can be
transformed into those which would be obtained with Chabrier and Kroupa IMFs reducing them
by a factor 0.55 and 0.62 respectively, without loosing precision beyond the limit of
20\%.
On the other hand, the proposed fit cannot be considered
a reliable mass estimator given the {\it a priori} assumptions made to define it, 
such as the fixed solar metallicity, a fixed very simple star formation history, a fixed
formation redshift that fixes a single possible age for the stellar populations of
the galaxies at a given redshift.

\vskip 0.2truecm
The second part of our work is dedicated to quantify the differences which can result
in the mass and luminosity estimates obtained by means of the best fitting technique
applied to the photometric SEDs of early-type galaxies at $1<z<2$. The main difference
with the previous part is due to the fact that age is not fixed {\it a priori} and
the analysis of the observed galaxies consists just in the determination of those
model parameters (such as age, star formation time scales, metallicity) which
produce the template best fitting the photometric data. 
In order to asses the reliability of this technique in deriving the luminosity
and stellar masses of galaxies, we applied it to a set of simulated galaxies for 
which the input properties were known. Mock photometric catalogs reproducing
the wavelength coverage and photometric accuracy of two main current surveys (i.e. VVDS and GOODS)
 have been built on the basis of models of early-type galaxies at $1<z<2$ assuming
BC03 code, solar metallicity and Salpeter IMF. Templates include also models
of galaxies which experienced a secondary star forming episode considering
different combinations of strengths and ages of the burst.
Then, we applied the best fitting technique to the mock catalogs by means of
{\it hyperz} and assuming the typical parameters used to analyze the early-type galaxies.
Once a best fitting template has been associated to each simulated galaxy, 
we compared the results obtained in the luminosity and stellar mass content estimates
with the true known input values. As far as the luminosities are concerned, we found that for the same goodness
of the best fit selected to represent the observations, the uncertainties in the determination
of the V absolute magnitude are much larger than in the K band.
In other words, it is easier to obtain a good determination
of near-IR absolute luminosities than optical ones. For what masses are concerned,
we found that the goodness of the mass estimate is  dependent on the mass estimator
adopted to derive it. The best mass estimate, not affected by any systematic trend, is
that derived by the scaling factor between templates and data on average, that
retrieves the correct value within a factor two. Alternatively,
the near-IR bands can be safely used to derive the mass content of early-type galaxies, while
optical bands produce much larger uncertainties and average underestimates of the stellar
masses which are difficult to be taken into account.

\vskip 0.2truecm
The uncertainties mentioned above refer to a favorable and not realistic situation
in which the SEDs of the observed galaxies can be well reproduced by means of the templates used
to find the best fitting model. Indeed, both model galaxies and fitting templates are
derived by means of BC03 code at solar metallicity and assuming Salpeter IMF.
In order to take into account the fact that synthetic templates are only approximation of 
the properties of real galaxies, we built an other set of simulated galaxies on the basis of templates including
different metallicities and based on other codes considered in the present work (i.e.
PEGASE, Ma05, GRASIL). All the galaxies are modeled assuming Salpeter IMF and following
the same combinations of ages and redshift of the previous case.
We used this new enlarged set of early-type galaxy models at $1<z<2$
to build  new mock photometric catalogs, assuming again the uncertainties of the GOODS
and VVDS surveys.
The analysis of this new mock catalogs has been performed in the same way of the previously
built ones. In particular, we continued to adopt a set of templates
at solar metallicity and based on the BC03 code for the comparison in {\it hyperz}.
{\bf We find that the true mass content of the galaxies cannot be retrieved
better than within a factor of 3}, being the only difference among the different estimators their
trend to over or under estimate it on average.

\vskip 0.2truecm
As a final test useful to answer to the question weather different models based
on different codes can bring significantly different results in the mass
estimate of early-type galaxies, we applied the best fitting technique
to a set of 10 massive early-type galaxies at $z\approx1.5$ 
(Saracco et al. 2003; Saracco et al. 2005; Longhetti et al. 2005).
We adopted 4 sets of templates built on the basis of the spectrophotometric codes BC03, Ma05, PEGASE and GRASIL,
among which we separately found the one best fitting the observed photometric SED of the galaxies.
We found that stellar mass estimates 
obtained with the Ma05 set of templates are on average $0.8\pm0.2$ than those obtained with the BC03 set,
while in case of young (i.e. $z>1.5$) galaxies the same ratio is 0.6.
Stellar mass estimates obtained with the PEGASE set of templates
are on average larger by a factor 1.1 than those obtained with
the BC03 set of templates, while those obtained with the GRASIL set of templates
result on average in agreement with those found with the BC03 set.

Summarizing, the use of different codes, at least those listed and discussed here,
does not bring any significant change in the final results when stellar mass
content of early-type galaxies are calculated by means of best fitting their
spectrophotometric data. In particular, when masses are estimated by means of 
comparison with the BC03 set of templates, the obtained values are on average between 0.8-1.3
of those which could be obtained with the other sets of templates here considered. 
On the other hand, the intrinsic uncertainty in the stellar mass estimate 
is larger than the former range, being not better than a factor 3
as previously stated.

\vskip 0.2truecm
Finally, we proposed a new {\it empirical} mass estimator
\mas$_{K}^{(V-K)}$ defined on the basis of a {\it raw} measure of the K band luminosity (M$_{Kraw}$)
and of the \mastol\ ratio in the K band calculated as simple function of
the {\it apparent} (V-K) colour (\mastol$_{K}^{(V-K)}$). Both the two involved quantities 
M$_{Kraw}$ and \mastol$_{K}^{(V-K)}$
do not need any multi-wavelength data fitting. Indeed, 
M$_{Kraw}$ is the absolute K band magnitude derived as
$m-d\_mod(z)-kcor_{raw}$, where $kcor_{raw}$ is the value calculated by means
of the fitting functions supplied in the first part of the work and depending only
on the value of $z$. The value of \mastol$_{K}^{(V-K)}$ is obtained from the apparent
(V-K) colour as:
$$\log(\mathcal{M/L}_{K}^{(V-K)})=0.086\times(V-K)_{obs}-0.863.$$
In spite of its simplicity, \mas$_{K}^{(V-K)}$ allows to obtain mass
estimates even more precise than \mas$_{b}$ (i.e. within a factor 2-2.5).
Furthermore, contrary to the mass
estimators based on the fitting of photometric data, \mas$_{K}^{(V-K)}$ is only slightly
dependent on the available photometric quality.
Its simplicity  and its stability with respect
to photometric uncertainties make the new mass estimator \mas$_{K}^{(V-K)}$
useful to derive stellar mass estimates of 
samples of {\it bona fide} elliptical galaxies at $1<z<2$ 
(e.g., to the samples of EROs with morphological information). The main advantage
in the use of  \mas$_{K}^{(V-K)}$ as stellar mass estimator is its independence
of models, because its calibrations has been proved over a wide range of models
and model parameters. This allows to compare masses estimated on different samples
in an umbiased way and with no need of any {\it a priori} correction or conversion
among different reference models.

\section*{Acknowledgments}
We want to thank the anonymous referee for all her/his helpful comments that
greatly improved the paper.
This research has received financial support from the
Istituto Nazionale di Astrofisica (Prin-INAF CRA2006 1.06.08.04).

\appendix
\section{Mass to light ratios from the best fitting technique}

\begin{figure}
\centering
\includegraphics[width=8.5truecm]{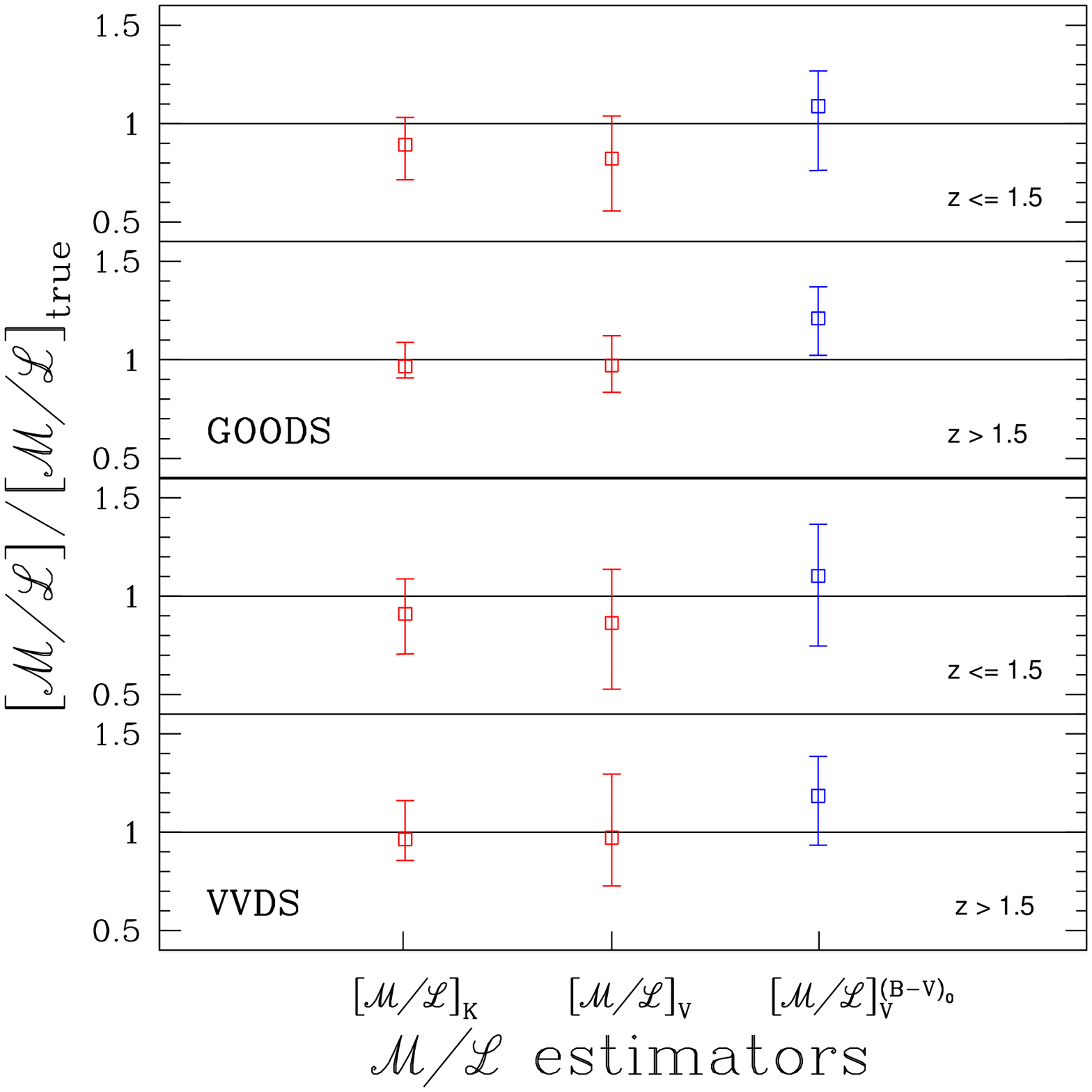}
\caption{Retrieved mass to light ratios \mastol\  in the K and V band
are reported as normalized to their true values [\mastol]$^{true}$.
Open square points represent the median values of the ratio between
the retrieved mass value and the true one. The errorbars represent the
16th (lower bar) and 84th (upper bar) percentiles of the distribution of
the same ratios, that
means that the 68\% of the results lie within the errorbars.
The upper two panels summarize the results obtained with the GOODS mock catalog,
while the lower two ones those obtained with the VVDS mock catalog.
Results are presented for simulated galaxies at $z\le1.5$ (upper first
and third panels) and at $z>1.5$.
}
\end{figure}

In \S4 we presented the results found by comparing the luminosity and
mass estimates obtained by the best fitting technique with the known input
values used to simulate some mock catalogs.
Here, we briefly analyze the same comparison applied to the values of \mastol\
ratios in both the K and V bands.
We derive
\mastol$_{K}$ and \mastol$_{V}$ which are the mass to light ratios
corresponding to the best fitting template of each simulated galaxy. We also calculated
$[\mathcal{M/L}_{V}]_{(B-V)_{0}}$ that is the mass to light ratio in the V band derived
from the rest frame (B-V)$_{0}$ colour following the Bell et al. (2005) prescription:
$$log([\mathcal{M/L}_{V}]^{(B-V)_{0}})=-0.628+1.305 (B-V)_{0}$$ 
The rest frame (B-V)$_{0}$ colour has been taken from the best fitting template of each
galaxy.

In figure A1 the \mastol\  ratios in the K and V band defined as detailed above are reported
normalized to their true values. 
The retrieved values of \mastol$_{K}$ are well within 0.7-1.1 (0.5-1.8) from the true values 
for 68\% (100\%) of the simulated galaxies.
At lower redshift for both the two catalogs it can be noted a trend to
underestimate the true value of \mastol$_{K}$ by a factor of about 0.9, while
at higher redshift (i.e. for the subsample of galaxies at only young ages) \mastol$_{K}$ retrieves the correct value
within 0.05.  
The underestimate of the \mastol\ 
value in the V band and at $z\le1.5$ is even larger (i.e. around 0.8), and the uncertainties are 
much larger in this band than in the K band. Indeed, \mastol$_{V}$ retrieves
the correct value within 0.5-1.3 (0.2-2.4) for 68\% (100\%) of the simulated galaxies. 
The values of [\mastol$_{V}$]$^{(B-V)_{0}}$ retrieved by means of the formula by Bell et al. (2005) 
show a smaller range of uncertainties with respect to \mastol$_{V}$ obtained from
the best fitting templates, but it displays a large trend to overestimate the true value of \mastol$_{V}$
(i.e., by a factor 1.1-1.2).

\label{lastpage}


\begin{thebibliography}{}
\bibitem[]{} Allen C. W., 1973, Astrophysical Quantities, Athlon Press, London
\bibitem[]{} Alongi M., Bertelli G., Bressan A., Chiosi C., Fagotto F., Greggio L., Nasi E. 1993, 
A\&AS 97,851
\bibitem[]{} Bell E.F., Papovich C., Wolf C., et al. 2005, ApJ 625, 23
\bibitem[]{} Bell E.F., Naab T., McIntosh D.H., et al. 2006, ApJ 640, 241
\bibitem[]{} Bender R., Saglia R.P., Ziegler B., Belloni P., Greggio L., Hopp U., Bruzual G. 1998,
ApJ 493, 529
\bibitem[]{} Bolzonella M., Miralles J.-M. \& Pell\`o R. 2000, A\&A 363, 476
\bibitem[]{} Bournaud F., Jog C. J. \& Combes, F. 2007, A\&A 476, 1179 
\bibitem[]{} Bressan A., Fagotto F., Bertelli G., Chiosi, C. 1993, A\&AS 100, 647
\bibitem[]{} Bruzual A.,G. \& Charlot S. 2003, MNRAS 344, 1000 (BC03)
\bibitem[]{} Calzetti D., Armus L., Bohlin R. C., Kinney A. L., Koorneef J., Storchi-Bergmann R. 2000, ApJ 533, 68
\bibitem[]{} Cassisi S. \& Salaris, M. 1997, MNRAS 285, 593
\bibitem[]{} Cassisi S., degl'Innocenti S., Salaris, M. 1997, MNRAS 290, 515
\bibitem[]{} Cassisi S., Castellani V., Ciarcelluti P., Piotto G., Zoccali M. 2000, MNRAS 315, 679
\bibitem[]{} Chabrier G. 2003, PASP 115, 763
\bibitem[]{} Charlot S. \& Bruzual G. 2008, in preparation
\bibitem[]{} Cimatti, A., Cassata P., Pozzetti L., et al. 2008, in press [astro-ph0001.1184]
\bibitem[]{} Clegg R.E.S. \& Middlemass D. 1987, MNRAS 228, 759
\bibitem[]{} Daddi E., Renzini A., Pirzkal N., et al. 2005, ApJ, 626, 680
\bibitem[]{} di Serego Alighieri S.,Vernet J., Cimatti A., et al. 2005, A\&A, 442, 125
\bibitem[]{} Elsner F., Feulner G. \& Hopp U. 2008, A\&A 477, 503
\bibitem[]{} Faber, S.M. \& Jackson R.E. 1976, ApJ 204, 668
\bibitem[]{} Fagotto F., Bressan A., Bertelli G., Chiosi C. 1994a, A\&AS 104, 365
\bibitem[]{} Fagotto F., Bressan A., Bertelli G., Chiosi C. 1994b, A\&AS 105, 29
\bibitem[]{} Fioc, M \& Rocca-Volmerange, B. 1997, A\&A 326, 950 (PEGASE)
\bibitem[]{} Girardi L., Bressan A., Chiosi C., Bertelli G., Nasi E. 1996, A\&AS 117, 113
\bibitem[]{} Grazian A., Fontana A., de Santis C., et al. 2006, A\&A 449, 951
\bibitem[]{} Groenewegen M.A.T. \& de Jong T. 1993, A\&A 267, 410
\bibitem[]{} Kroupa P., 2001, MNRAS, 322, 231
\bibitem[]{} Kurucz R.L. 1992, IAUS 149, 225
\bibitem[]{} Lancon A. \& Mouhcine M. 2002, A\&A 393, 167
\bibitem[]{} Le Fevre O., Guzzo L., Meneux B., et al. 2005, A\&A 439, 845
\bibitem[]{} Longhetti M., Saracco P., Severgnini P., et al. 2005, MNRAS 361, 897
\bibitem[]{} Maraston C. 1998, MNRAS 300, 872
\bibitem[]{} Maraston C. 2005, MNRAS 362, 799 (Ma05)
\bibitem[]{} Maraston C., Daddi E., Renzini A., et al. 2006, ApJ 625, 85
\bibitem[]{} Marigo P. \& Girardi L. 2007, A\&A 469, 239
\bibitem[]{} McCarthy P.J., Le Borgne D., Crampton D., et al. 2004,  ApJ 614, L9
\bibitem[]{} Mignoli M., Cimatti A., Zamorani G., et al. 2005, A\&A 437, 883
\bibitem[]{} Rauch T. 2003, A\&A 403, 709
\bibitem[]{} Renzini A. 2006, ARA\&A 44, 141
\bibitem[]{} Rettura A. et al. 2006, A\&A 458, 717
\bibitem[]{} Salpeter E. E. 1955, ApJ 121, 161
\bibitem[]{} Saracco P., Longhetti M., Severgnini P., et al. 2003, A\&A 398, 127
\bibitem[]{} Saracco P., Longhetti M., Severgnini P., et al. 2005, MNRASLet 357, 40
\bibitem[]{} Saracco P., Longhetti M., Andreon S. 2008, MNRAS {\it in press, arXiv:0810.2795}
\bibitem[]{} Shen S., Mo H.J., White S., Blanton M.R., Kauffmann G., Voges W., Brinkmann J., Csabai I. 2003,
MNRAS 343, 978
\bibitem[]{} Silva, L., Granato, G.L., Bressan, A., Danese, L., 1998, ApJ 509, 103 (GRASIL)
\bibitem[]{} Thomas, D., Maraston, C., Bender, R., Mendes de Oliveira, C. 2005, ApJ 621, 673
\bibitem[]{} Treu T., Ellis R.S., Liao T.X., al. 2005, ApJ, 633, 174
\bibitem[]{} van der Wel A., Franx M., van Dokkum P. G., Rix H.-W. 2004
ApJ, 601, L5
\bibitem[]{} van der Wel A., Franx M., van Dokkum P. G., Rix H.-W., Illingworth G. D. \&  Rosati, P.
2005, ApJ 631, 145
\bibitem[]{} van Dokkum P. G., AJ 130, 2647
\bibitem[]{} van Dokkum P. G., Franx M., Kelson D. D., Illingworth G. D., Fisher D., Fabricant D. 1998, ApJ 500, 714
\bibitem{}{} Vanzella E., Cristiani S., Dickinson M., {\it et al.} 2008, A\&A 478, 83
\bibitem[]{} Vassiliadis E. \& Wood P.R 1993, ApJ 413, 641
\bibitem[]{} Ziegler B.L., Thomas D., Böhm A., Bender R., Fritz A., Maraston C. 2005, A\&A 433, 519

\end{thebibliography}
\end{document}